\documentclass{aa}  

\usepackage{graphicx}
\usepackage{caption}
\usepackage{subcaption}
\usepackage{xcolor}
\usepackage{txfonts}
\newcommand{\kms}{\ensuremath{\rm km\,s^{-1}}\xspace}
\newcommand{\Msun}{\ensuremath{\rm M_{\sun}}\xspace}
\newcommand{\Lsun}{\ensuremath{\rm L_{\sun}}\xspace}
\newcommand{\Jyb}{\ensuremath{\rm Jy\,beam^{-1}}\xspace}
\newcommand{\commentstere}[1]{\textcolor{black}{#1}}

\begin{document}

   \title{PRODIGE - Envelope to disk with NOEMA \thanks{Based on observations carried out under project number L19MB with the IRAM NOEMA Interferometer. IRAM is supported by INSU/CNRS (France), MPG (Germany) and IGN (Spain)}}

   \subtitle{I. \commentstere{A 3000 au} streamer feeding a Class I protostar}

   \author{M. T. Valdivia-Mena
          \inst{1}
          \and
          J. E. Pineda\inst{1}
          \and
          D. M. Segura-Cox\inst{2}\thanks{Astronomy and Astrophysics Postdoctoral Fellow}
          \and
          P. Caselli\inst{1}
          \and
          R. Neri\inst{3}
          \and 
          A. L\'opez-Sepulcre\inst{3,4}
          \and
          N. Cunningham\inst{4}
          \and
          L. Bouscasse\inst{3}
          \and
          D. Semenov\inst{5}
          \and
          Th. Henning\inst{5}
          \and
          V. Pi\'etu\inst{3}
          \and
          E. Chapillon\inst{6,3}
          \and
          A. Dutrey\inst{6}
          \and
          A. Fuente\inst{7}
          \and
          S. Guilloteau\inst{6}
          \and
          T. H. Hsieh\inst{1}
          \and
          I. Jim\'enez-Serra\inst{8}
          \and
          S. Marino\inst{5}
          \and
          M. J. Maureira\inst{1}
          \and
          G. V. Smirnov-Pinchukov\inst{5}
          \and
          M. Tafalla\inst{7}
          \and
          B. Zhao\inst{9}
          }

   \institute{Max-Planck-Institut f\"ur extraterrestrische Physik, Giessenbachstrasse 1, D-85748 Garching, Germany\\
   \email{mvaldivi@mpe.mpg.de}
   \and
   Department of Astronomy, The University of Texas at Austin, 2515 Speedway, Austin, TX 78712, USA
   \and
    Institut de Radioastronomie Millim\'{e}trique (IRAM), 300 rue de la Piscine, F-38406, Saint-Martin d'H\`{e}res, France
    \and
    IPAG, Universit\'{e} Grenoble Alpes, CNRS, F-38000 Grenoble, France
    \and
    Max-Planck-Institut f\"{u}r Astronomie, K\"{o}nigstuhl 17, D-69117 Heidelberg, Germany
    \and
    Laboratoire d'Astrophysique de Bordeaux, Universit\'{e} de Bordeaux, CNRS, B18N, All\'{e}e Geoffroy Saint-Hilaire, F-33615 Pessac, France
    \and
    Observatorio Astron\'{o}mico Nacional (IGN), Alfonso XII 3, 28014, Madrid, Spain
    \and
    Centro de Astrobiolog\'{i}a (CSIC-INTA), Ctra. Ajalvir km 4, Torrej\'{o}n de Ardoz, E-28850, Madrid, Spain
    \and
    Department of Physics and Astronomy, McMaster University, Hamilton, ON L8S 4E8, Canada
    }

   \date{\today}

 
  \abstract
   {In the past few years, there has been a rise in the detection of streamers, asymmetric flows of material directed toward the protostellar disk with material from outside the star's natal core. It is unclear how they affect the process of mass accretion, in particular beyond the Class 0 phase.}
   {We investigate the gas kinematics around Per-emb-50, a Class I source in the crowded star-forming region NGC 1333. Our goal is to study how the mass infall proceeds from envelope to disk scales in this source.}
   {We use new NOEMA 1.3 mm observations, including C$^{18}$O, H$_2$CO and SO, in the context of the PRODIGE MPG - IRAM program, to probe the core and envelope structures toward Per-emb-50.}
   {We discover a streamer delivering material toward Per-emb-50 in H$_2$CO and C$^{18}$O emission. The streamer’s emission can be well described by the analytic solutions for an infalling parcel of gas along a streamline with conserved angular momentum, both in the image plane and along the line of sight velocities. The streamer has a mean infall rate of $1.3 \times 10^{-6}$ \Msun yr$^{-1}$, $5-10$ times higher than the current accretion rate of the protostar. SO and SO$_2$ emission reveal asymmetric infall motions in the inner envelope, additional to the streamer around Per-emb-50. Furthermore, the presence of SO$_2$ could mark the impact zone of the infalling material.  }
   {The streamer delivers sufficient mass to sustain the protostellar accretion rate and might produce an accretion burst, which would explain the protostar’s high luminosity with respect to other Class I sources. Our results highlight the importance of late infall for protostellar evolution: streamers might provide a significant amount of mass for stellar accretion after the Class 0 phase.}


   \maketitle
%


\section{Introduction}

The classical picture of star formation allows us to understand the collapse of a dense, individual core through simple physical assumptions, but does not fully explain the current observations of protostars and protoplanetary disks.
In general, the classical models consist of a dense, mostly isolated core inside a molecular cloud which undergoes axisymmetric collapse and, due to the conservation of angular momentum, flattens and creates a disk around the central protostar \citep[e.g.,][]{Shu1977corecollapse, Terebey1984rotation}. 
The first limitation of the classical models is that they depend on two assumptions: the spherical symmetry of the core collapse and its lack of interaction with material outside the protostar's natal core.
In reality, molecular clouds are asymmetric 
at all scales \commentstere{ \citep{Andre2014PPVIFilaments, Pineda2022}}, from the parsec-sized filaments \citep[e.g., ][]{HacarTafalla2011Taurus, Andre2010GBSHerschel}, 
to asymmetric envelopes around protostars \citep{Tobin2010Class0Envelopes}.
Numerical simulations of molecular clouds that follow the collapse of several cores, including turbulence and magnetic fields, can reproduce these observed filaments and asymmetric structures \citep[e.g., ][]{Lebreuilly2021MFieldDiskForm,Kuznetsova2019, Kuffmeier2017infalltodisks,Padoan2014Infallsim}.

A second problem with the standard model of inside-out, axisymmetric collapse of an isolated core is that predicts a constant mass accretion rate $\sim 10^{-5}$ \Msun yr$^{-1}$ \citep{Stahler1980ProtoEvolutionI}, but observed bolometric luminosities in embedded protostars imply accretion rates that are 10 to 100 times lower than this value \citep{Kenyon1990lumproblem, Evans2009C2Dlifetime}. This is known as the ``luminosity problem". Proposed solutions to this problem include an initial strong accretion phase followed by an accretion rate decay \citep{Padoan2014Infallsim}, and strong bursts of accretion during the protostellar phase \citep{Kuffmeier2018EpisodicAcc, Zhao2018_3Ddec_mag_field, Vorobyov2015DiskGI}. These solutions show that \commentstere{the} accretion process is asymmetric both in space and time, which is incompatible with fully axisymmetric collapse. 
Therefore, even if the simple symmetric model allows for a comprehension of isolated sources, it does not capture all the phenomena that affect the star formation process.


Recently, numerical simulations show that the local environment surrounding the protostar has a deep impact on its evolution \citep{Hennebelle2020diskformationsims,Kuffmeier2018EpisodicAcc,Kuffmeier2017infalltodisks, Padoan2014Infallsim}. In particular, simulations focusing on star and disk formation repeatedly find asymmetric flows toward the disk \citep[e.g.,][]{Wurster2019, Kuznetsova2019,Kuffmeier2019-Bridge,Kuffmeier2017infalltodisks}. 
These long, thin inflows, called streamers, can deliver mass from outside the natal core to increase the available mass for the protostar \citep{Pelkonen2021massbeyondcore} and might have effects in the structure of protoplanetary disks \citep{Kuffmeier2017infalltodisks}.
All these simulations show that the collapse from core to protostar is more complex than axisymmetric inside-out collapse. 

In the last few years, observations started to find streamers from envelope to disk scales \commentstere{ \citep[see][and references within]{Pineda2022}}. 
Streamers are found from the highly embedded Class 0 phase \citep{Pineda2020,LeGouellec2019PolarizedClass0} through the less embedded Class I \citep[Segura-Cox et al. in prep.,][]{Chou2016DiskandFilconnectionL1455}, all the way 
to Class II sources \citep[e.g.,][]{Ginski2021,Garufi2021arXiv-accretionDGTauHLTau,Alves2020,Akiyama2019,Yen2019HLTau,Tang2012ABAurLateAccretion}. They have also been found feeding not only singles, but also protostellar binaries, both funneling material toward the inner circumstellar disks \citep{Phuong2020GGTauSpirals,Alves2019PretzelAccretion, Dutrey2014GGTau} and to the binary system as a whole \citep{Pineda2020}.
These structures are observed in a diversity of molecules, such as $^{12}$CO \citep{Alves2020} 
and HC$_3$N \citep{Pineda2020}, and also in scattered light \citep{Ginski2021,Akiyama2019}. 
The first streamer to be characterized using only free-fall motion, and thus confirming it is infalling toward the protostar, is located toward the Class 0 source Per-emb-2 \citep{Pineda2020}. This streamer transports material from outside the dense core ($> 10\,000$ au) into the protoplanetary disk and protostar system. The infall rate of this streamer, which describes how much mass is deposited into disk forming scales, is comparable to the accretion rate toward the protostar, implying that the streamer could change the future protostellar accretion rate by funneling extra material. This streamer was discovered with a carbon-chain species, HC$_3$N, which traces best the less chemically evolved material in contrast to the more evolved protostellar core seen in N$_2$H$^+$ \citep{BerginTafalla2007coldcloudsreview}. 
These objects prove that the environment influences the star’s development and support the results from simulations that state the mass available to the protostar could be coming from further away than the natal core \citep{Pelkonen2021massbeyondcore}.


Even though asymmetric infall is an ubiquitous feature in numerical simulations, to the best of our knowledge, only a few streamers have been found and their infall properties quantified using either average estimates of infalling material and/or free-fall motion models toward the disk and protostar system \citep[e.g.,][]{Ginski2021,Pineda2020,Alves2019PretzelAccretion}. 
This is where the MPG - IRAM observing program ``PROtostars \& DIsks: Global Evolution" (PRODIGE, CO-PIs: P. Caselli, Th. Henning) comes in: this program is designed as a coherent study of the physical and chemical properties of young protostellar systems, targeting 32 Class 0/I sources and 8 Class II T Tauri protoplanetary disks. 
One of its goals is to search for material flowing into Class 0 and I sources and investigate the mass budget during these phases. 
PRODIGE observations are done with the IRAM NOrthern Extended Millimetre Array (NOEMA), located at the Plateau de Bure in the French Alps. 
This program takes advantage of the PolyFix correlator to make an unbiased survey of molecular lines, thus allowing for the search of streamers in multiple chemical tracers.  

In this paper, we present new NOEMA 1.3 mm ($\approx220$ GHz) observations from the PRODIGE survey of 5 molecules (H$_2$CO, C$^{18}$O, $^{12}$CO, SO and SO$_2$) toward the Class I protostar Per-emb-50. Our aim is to characterize the core kinematics around this embedded protostar from approximately 300 au scales out to 3000 au from the source, to investigate how the mass infall proceeds from envelope to disk scales.
The paper is divided as follows. 
Section \ref{sec:observations} describes 
the NOEMA observations, data reduction and imaging procedures.
Section \ref{sec:results} shows the observed structures in each molecular tracer and how we separate the different kinematic components. 
We discuss how the structures found might affect the protostar and protostellar disk evolution in Per-emb-50 and how they fit in the general star formation paradigm in Sect. \ref{sec:discussion}. 
We summarize our results in Sect. \ref{sec:conclusions}.

\section{Observations and data reduction\label{sec:observations}}

\subsection{Per-emb-50}


Per-emb-50 is an embedded Class I protostar, according to its Spectral Energy Distrubution (SED) in the near- and mid-infrared \citep{Evans2009C2Dlifetime, Enoch2009}. It is located in the active star-forming region NGC 1333, at a distance of 293 pc \citep{Ortiz-Leon2018,Zucker2018distance}, in the Perseus Giant Molecular Cloud. This protostar is $\sim10$ times brighter than other Class I sources in the vicinity \citep{Dunham2015gouldbeltcatalog,Enoch2009} and its protostellar accretion rate is estimated between $(1.3-2.8) \times 10^{-7}$ \Msun yr$^{-1}$, also around 10 times larger than other Class I sources \citep{Fiorellino2021-mdot}. It has a clear outflow observed in $^{12}$CO\,(2--1) emission with an east-west orientation \citep{Stephens2019}. 

VLA 8 mm continuum analysis shows a large dust disk in Per-emb-50, with a characteristic radius between $27-32$ au (where there is a significant drop in the dust flux profile), and dust mass around $0.28-0.58$ \Msun \citep{Segura-Cox2016}. Radiative transfer models applied to millimeter observations suggest that grain growth has proceeded within the envelope, producing grains with sizes $\sim100$ $\mu$m \citep{Agurto-Gangas2019}.

Properties of the protostar and its disk taken from the literature are summarized in Table \ref{tab:peremb50}.

\begin{table}[htb]
\caption{Properties of Per-emb-50 from literature. }
\centering
\begin{tabular}{ccc}
\hline\hline   
Property     & Value  & Reference \\
\hline
RA (J2000, deg)    & 03:29:07.76 & 1 \\
Dec (J2000, deg)     & +31:21:57.2 & 1 \\
$M_{*}$ (\Msun) & $1.5-1.9$   & 2 \\
$M_{disk}$ (\Msun) & $0.28 - 0.58$ & 3 \\
$R_{c}$ (au) $^{*}$ & $27 - 32$ & 3 \\
$i_{disk}$ ($\deg$)      & 67  & 3           \\
PA$_{disk}$ ($\deg$)     & 170   & 3  \\ 
$d$ (pc) $^{**}$    & $293 \pm 22$  & 4   \\ 
\hline
\end{tabular}
\tablefoot{
\tablefoottext{*}{This is a characteristic radius at which there is a significant drop in the dust flux exponential profile, a proxy for the radius.} 
\tablefoottext{**}{This distance corresponds to the distance to NGC 1333.}\tablefoottext{1}{\cite{Enoch2009}} \tablefoottext{2}{\cite{Fiorellino2021-mdot}} \tablefoottext{3}{\cite{Segura-Cox2016}}  \tablefoottext{4}{\cite{Ortiz-Leon2018, Zucker2018distance}}

\label{tab:peremb50}}
\end{table}

\subsection{NOEMA observations}

The observations are obtained with NOEMA and are part of the MPG-IRAM Observing Program PRODIGE (Project ID L19MB). In this program, we use the Band 3 receiver and the new PolyFix correlator, tuned with a local-oscillator (LO) frequency of 226.5 GHz. PolyFix provides a full 16 GHz of bandwidth at coarse spectral resolution (2 MHz channel width) and is divided into 4 units (LSB Outer, LSB Inner, USB Inner and USB Outer). Simultaneously, we place 39 high spectral resolution windows of 62.5 kHz channel resolution within the coarse resolution 16 GHz bandwidth.  

Observations of Per-emb-50 were conducted in two separate periods for each antenna configuration. The C configuration data were observed during 2019 December 29th and 2021 January 5th. The D configuration observations were taken in 2020 August 6th and September 7th and 8th. The maximum recoverable scale (MRS) for our data is 16.9\arcsec at 220 GHz, approximately 5000 au at the distance of Per-emb-50.

\begin{table*}[htbp]
\centering
\caption{\label{tab:cubeprops}Properties of the molecular line observations from NOEMA. }
\begin{tabular}{cccccccc}
\hline\hline   
Molecule & Transition & $E_{up}$  &  Frequency$^{*}$ & $\theta_{maj}\times \theta_{min}$ (PA)  & rms & $\Delta V_{\mathrm{chan}}$ \\
 & & (K) &(GHz)  &  ($\arcsec \times \arcsec$, $^{\circ}$)  & (mJy beam$^{-1}$) & (\kms)\\
\hline
SO        & 5$_5$ -- 4$_4$    & 44.1     & 215.2206530 & $1.25\times0.73$ (21.48) &  13.01 & 0.08 \\
H$_2$CO   & 3$_{0,3}$ -- 2$_{0,2}$ & 21.0  & 218.2221920 & $1.24\times0.72$ (21.43) & 11.97 & 0.08 \\
SO$_2$      & 11$_{1,11}$ -- 10$_{0,10}$     & 60.4    & 221.9652196 & $1.24\times0.72$ (20.89)  & 11.54 & 0.08 \\
C$^{18}$O & 2 -- 1       & 15.8    & 219.5603541 & $1.24\times0.71$ (20.87) & 13.94 & 0.08\\
$^{12}$CO & 2 -- 1     & 5.5    & 230.5380000 & $1.15\times0.67$ (21.20) & 7.43 & 2.60 \\
\hline
\end{tabular}
\tablefoot{
\tablefoottext{*}{Taken from the Cologne Database for Molecular Spectroscopy \citep{Endres2016CDMS}}
}
\end{table*}

We calibrate the data using the standard observatory pipeline in the GILDAS (Grenoble Image and Line Data Analysis Software) package CLIC (Continuum and Line Interferometer Calibration). We use 3C84 and 3C454.3 as bandpass calibrators. Phase and amplitude calibration sources are 0333+321 and 0322+222, and observations for these calibrators were taken every 20 min. LKHA101 and MWC349 are used as flux calibrators. The uncertainty in flux density is 10\%. The continuum is bright enough to allow for self calibration. Only for the continuum image used in this work, self calibration is performed iteratively with solution intervals of 300 s, 135 s, and 45 s. The line observations are not done with self-calibrated data. The resulting continuum image, shown in Appendix \ref{sec:cont}, is done with the Lower Inner (LI) continuum window and has a noise of 0.2 m\Jyb.

Continuum subtraction and data imaging are done with the GILDAS package \texttt{mapping} using the \texttt{uv\_baseline} and \texttt{clean} tasks. All line cubes are imaged using natural weight to minimize the rms, while the continuum maps are imaged with robust $=1$, to improve the angular resolution. We image the continuum-subtracted cubes using the standard CLEAN algorithm and a manual mask for each channel. Once we converge to a final mask, we perform a final CLEAN down to the rms level using multiscale CLEAN algorithm implemented in \texttt{mapping}. This has an effect of reducing imaging artifacts (mainly negative emission bowls), thus improving the general image quality around bright sources. 


Towards Per-emb-50, we detect $^{12}$CO\,(2--1), C$^{18}$O\,(2--1), H$_2$CO\,(3$_{0,3}$--2$_{0,2}$), SO\,($5_5$--$4_4$) and SO$_2$\,(11$_{1,11}$ -- 10$_{0,10}$) line emission. The $^{12}$CO\,(2--1) line is located in the coarse resolution bandwidth, whereas the rest of the lines are inside the high resolution windows. The final line cubes have a beam FWHM $\theta$ of approximately 1.2\arcsec, a primary beam FWHM of 22\arcsec at 220 GHz, a field of view (FoV) of 45.8\arcsec diameter and a channel spacing of 0.08 \kms. The effective spectral resolution is approximately 1.7 times the channel spacing. The average rms is around 13 m\Jyb or around 400 mK. The resulting properties of each line cube are reported in Table \ref{tab:cubeprops}. 

\section{Results\label{sec:results}}

\subsection{Streamer in Per-emb-50 \label{sec:asym-inflows}}


\begin{figure*}[!htbp]
     \centering
         \includegraphics[width=0.95\textwidth]{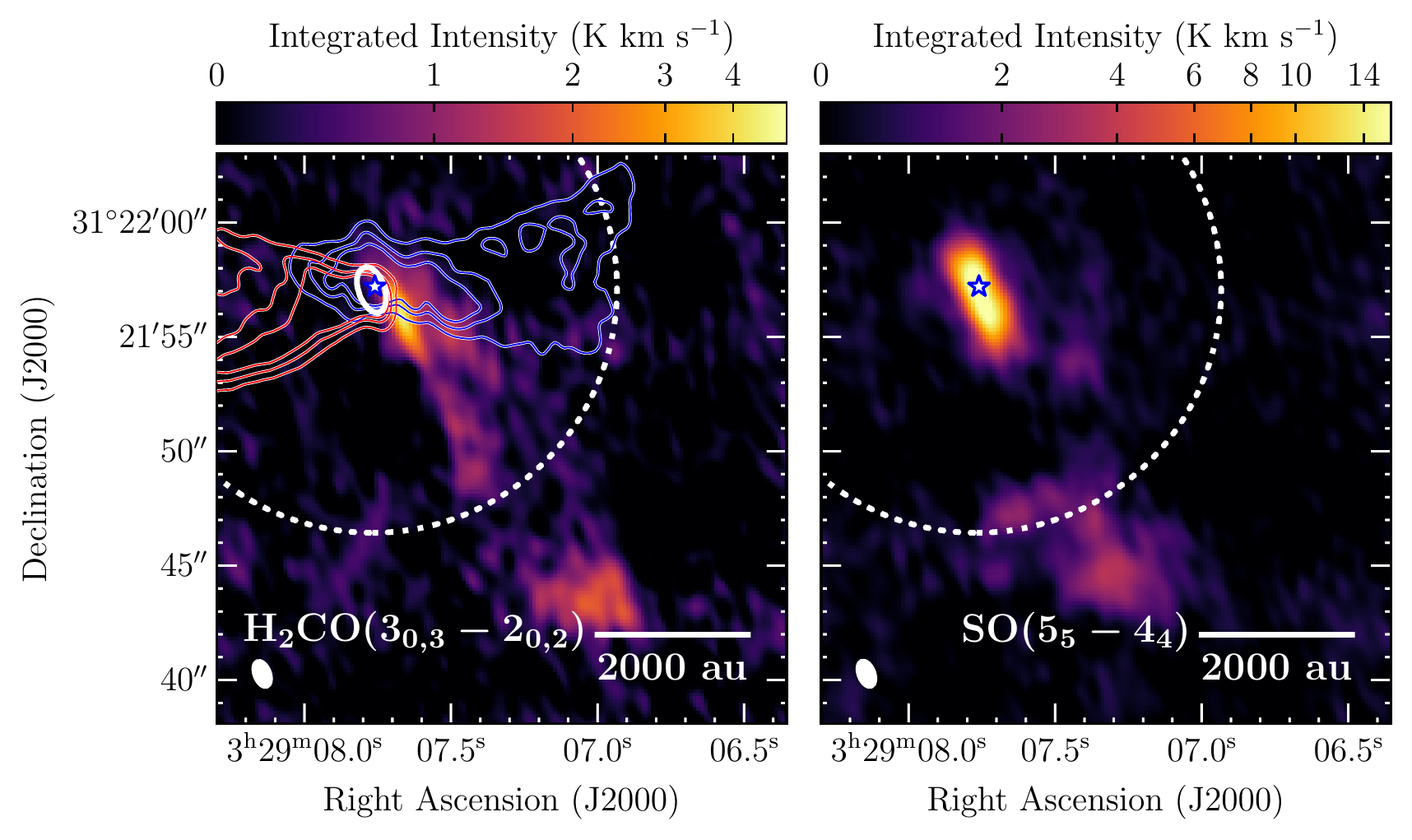}
        \caption{Integrated intensity images of H$_2$CO\,(3$_{0,3}$--2$_{0,2}$) and SO\,($5_5$--$4_4$) before primary beam correction are shown at the left and right, respectively. Primary beam FWHM sizes are represented with white \commentstere{filled} dashed circles in each image. The blue star represents the location of Per-emb-50. Beam sizes are represented by white ellipses in the bottom left corner of each image. \textbf{Left:} H$_2$CO\,(3$_{0,3}$--2$_{0,2}$) integrated intensity between 5.5 and 9.5 \kms. Red and blue contours correspond to the redshifted and blueshifted emisions coming from the outflow, respectively, traced in our wideband $^{12}$CO\,(2 -- 1) emission. Contour levels are shown at 8, 16 and 24 K \kms. The white contour represents the continuum emission at a 7 m\Jyb level (see Fig.\,\ref{fig:cont}). 
        \textbf{Right:} SO\,($5_5$--$4_4$) integrated intensity between -1 and 14 \kms.
        }
        
        \label{fig:images}
\end{figure*}

The integrated intensity images for H$_2$CO\,(3$_{0,3}$--2$_{0,2}$) and SO\,($5_5$--$4_4$) are shown in Figure \ref{fig:images}. The former unveils a large streamer in the south-west of the central star, whereas the latter shows extended emission surrounding the protostar. We refer to H$_2$CO(3$_{0,3}$--2$_{0,2}$) as H$_2$CO and SO($5_5$--$4_4$) as SO from now on. The integrated intensity maps are calculated between 5.5 and 9.5 \kms in the case of H$_2$CO and between -1 and 14 \kms in SO, which are the velocity ranges where all emission over $3\sigma$ in each channel (see Table \ref{tab:cubeprops} for $\sigma$ values) is present for each molecule.

The streamer stretches from the location of the protostar to the edge of the primary beam in southwest direction, with a total length of approximately 3000 au (22\arcsec) and a width of approximately 300 au (1\arcsec). As the width of the streamer is barely resolved, this width is considered an upper limit. Also, as the streamer reaches up to the primary beam FWHM, it is possible that it extends further, so the length is a lower limit as well. The peak integrated intensity in H$_2$CO presents a signal-to-noise ratio (S/N) of 11. The streamer is detected with an $\mathrm{S/N}\geq 6$ along its 3000 au length. 

This streamer is spatially unrelated to the outflow of Per-emb-50, since the emission does not spatially overlap with the outflow. Figure \ref{fig:images} Left shows the outflow observed in $^{12}$CO(2--1) emission, from the wide-band setup of our NOEMA observations. $^{12}$CO is integrated from $\mathrm{V}_{\mathrm{LSR}}=-4$ to 4 \kms for the blueshifted emission and from 11 to 20 \kms for the redshifted emission. The outflow, previously observed by \cite{Stephens2019}, is in the east-west direction, whereas the H$_2$CO streamer extends in a northeast-southwest direction. 


Outside the primary beam and to the southwest of Per-emb-50, there is also enhanced H$_2$CO and SO emission. 
\commentstere{It is difficult to characterize the nature of this structure because it is outside the primary beam, even after primary beam correction: emission in this region might be contaminated by emission from outside our field of view, leaking through the side-lobes of the antenna response pattern. }
In Section \ref{sec:discussion-streamlowlims}, we discuss the possibility that this emission consists of an extension of molecular emission further away from the protostar.



\subsection{Streamer kinematics\label{sec:H2COgausfit}}

\begin{figure*}[htbp]
     \centering
         \includegraphics[width=0.5\textwidth]{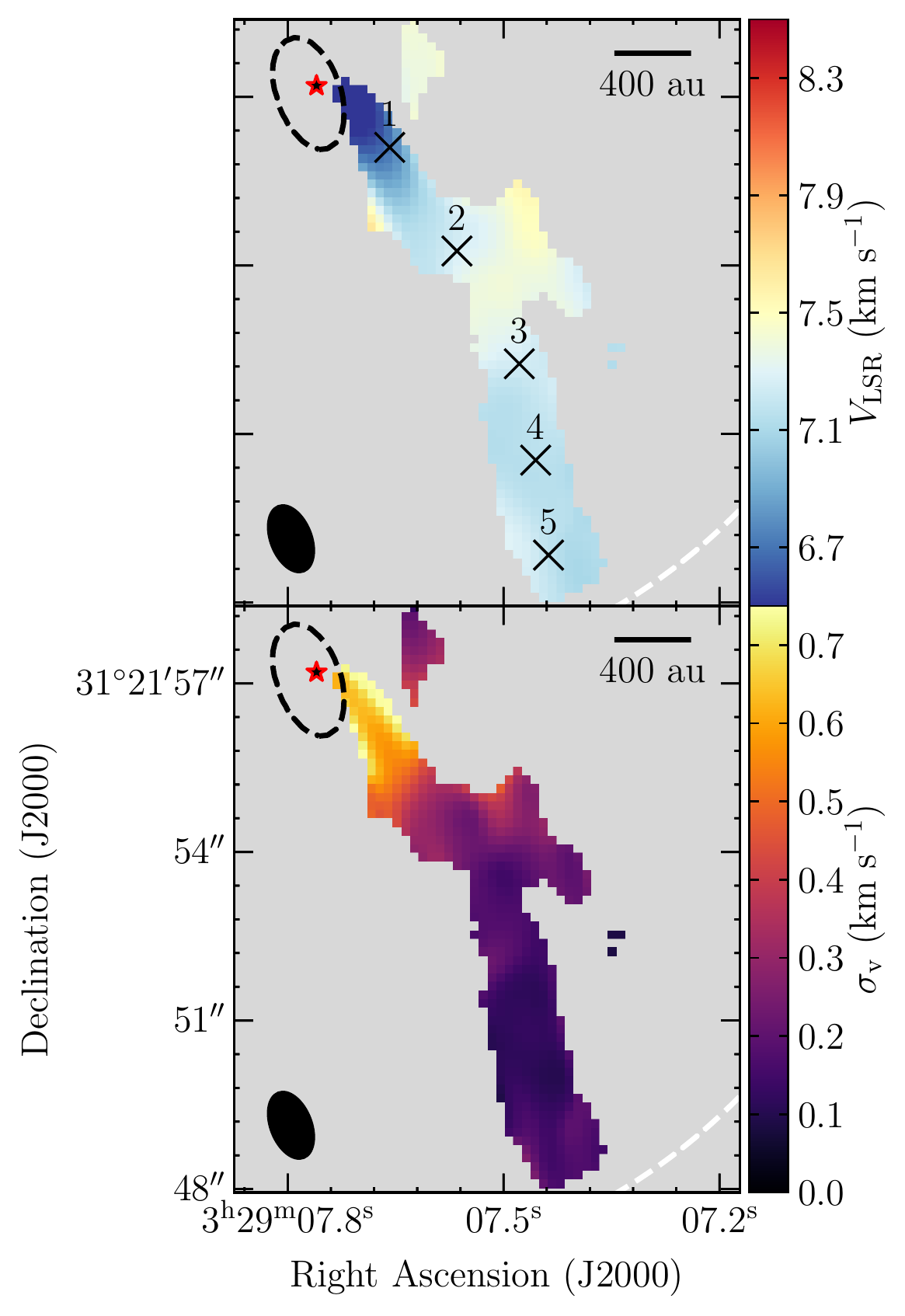}
         \includegraphics[width=0.4\textwidth]{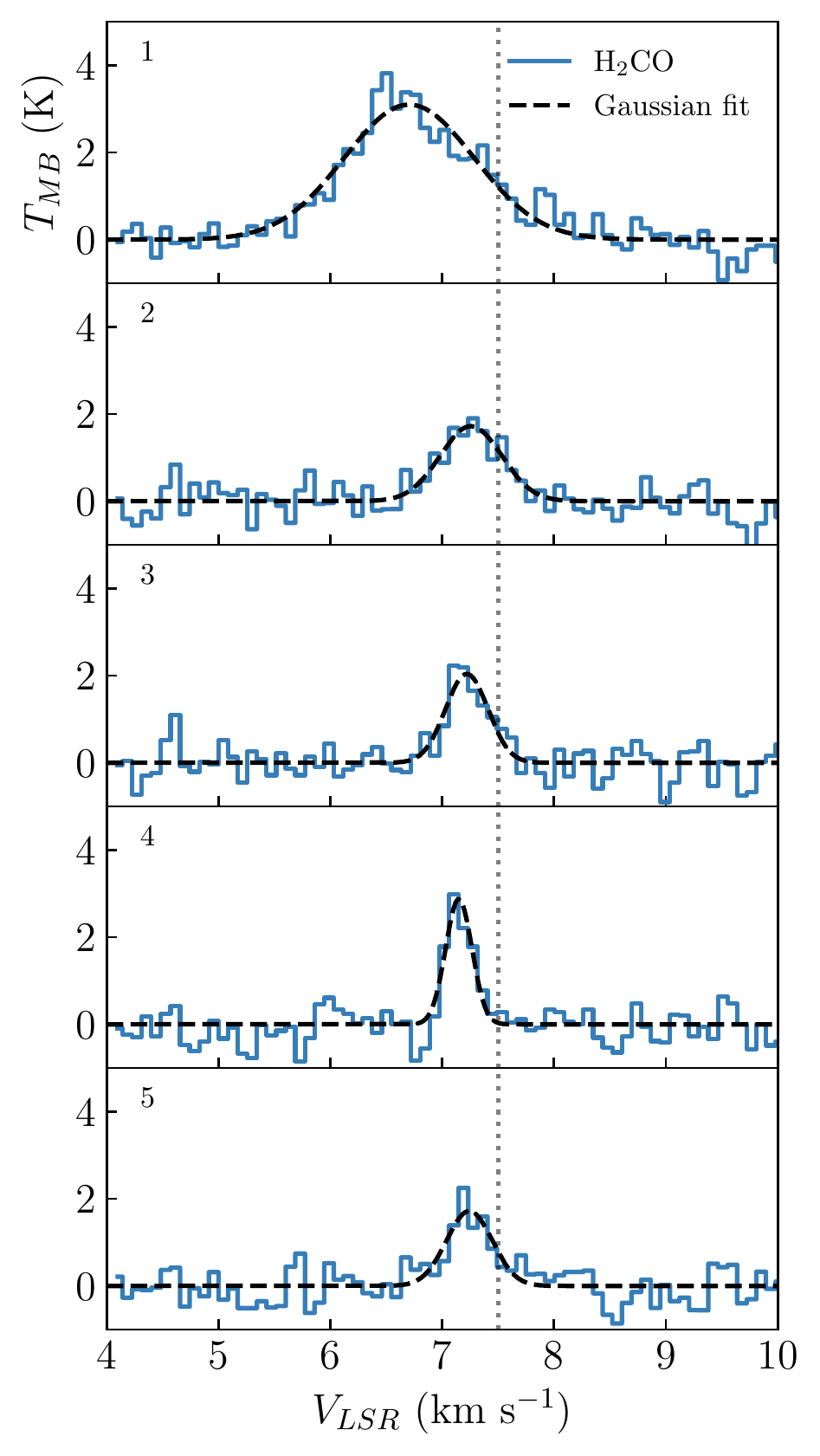}
        \caption{\textbf{Left: }Central velocity  $V_{\mathrm{LSR}}$ and velocity dispersion $\sigma_{\mathrm{v}}$ of the H$_2$CO streamer are shown in the top and bottom panel, respectively. These are obtained from the Gaussian model for H$_2$CO emission of each spectra with $\mathrm{S/N}>4$. The red star represents the central position of Per-emb-50. Black labeled crosses mark the positions where we extract spectra, shown to the right. Black dashed contours correspond to the continuum emission at a brightness level of 7 m\Jyb (see Fig.\,\ref{fig:cont}). White dashed lines represent the primary beam FWHM, centered at the location of Per-emb-50. The beam is drawn in the lower left corners of each image. \textbf{Right:} H$_2$CO spectra at selected positions along the streamer, together with the Gaussian that best fits each spectrum. Blue lines indicate the H$_2$CO spectra and the dashed black lines represent the best fit Gaussian function. The uncertainty in $T_{MB}$ is 0.3 K. The gray dotted line represents the protostar's $V_{LSR}=7.5$ \kms. } 
        \label{fig:H2COfit}
\end{figure*}

We fit a Gaussian to the H$_2$CO line emission without primary beam correction using \texttt{pyspeckit} (see Appendix \ref{ap:gaussfit} for details). The central velocity $V_{\mathrm{LSR}}$ and velocity dispersion $\sigma_{\mathrm{v}}$ of the Gaussians that best fit the spectrum \commentstere{at} each pixel with $\mathrm{S/N}>4$ are shown in Figure \ref{fig:H2COfit}. 

H$_2$CO line emission is characterized by mostly blueshifted emission with respect to Per-emb-50's $V_{\mathrm{LSR}}$ (7.5 \kms, see Section \ref{sec:protostellarmass}). The velocity of the streamer further away from the protostar consists of mostly constant blueshifted velocities (with $V_{\mathrm{LSR}}\approx7.2$ \kms) and low velocity dispersion of $\sigma_{\mathrm{v}}=0.1-0.2$ \kms. Closer to the protostar, between positions 2 and 3 marked in Fig.~\ref{fig:H2COfit} and shifted to the west with respect to the general direction of the streamer, there is a sudden increase in velocities, from 7.2 to 7.5 \kms. We refer to this region as the "kink" from now on, as it is a "kink" or bend in the overall shape of the emission and an abrupt break in the velocity distribution. It is improbable that the sudden redshift in velocities is caused by the outflow, as its west side consists of blueshifted emission, whereas the kink in the streamer is redshifted with respect to the rest of the streamer’s velocities. The kink is followed by a reversal back to \commentstere{blueshifted} velocities approaching the protostar, in the inner 1000 au. There is a steep velocity ($V_{\mathrm{LSR}}$) gradient, a change of 7.1 to 6.5 \kms in $\sim 750$ au, and the velocity dispersion ($\sigma_{\mathrm{v}}$) increases from  0.4 to 0.7 \kms in the same region. This gradient suggests that the gas follows infall motions dominated by the central gravitational force of the  protostar, disk and inner envelope. 



\subsection{Protostellar mass and velocity\label{sec:protostellarmass}}

\begin{figure*}[htbp]
     \centering
    \includegraphics[width=0.49\textwidth]{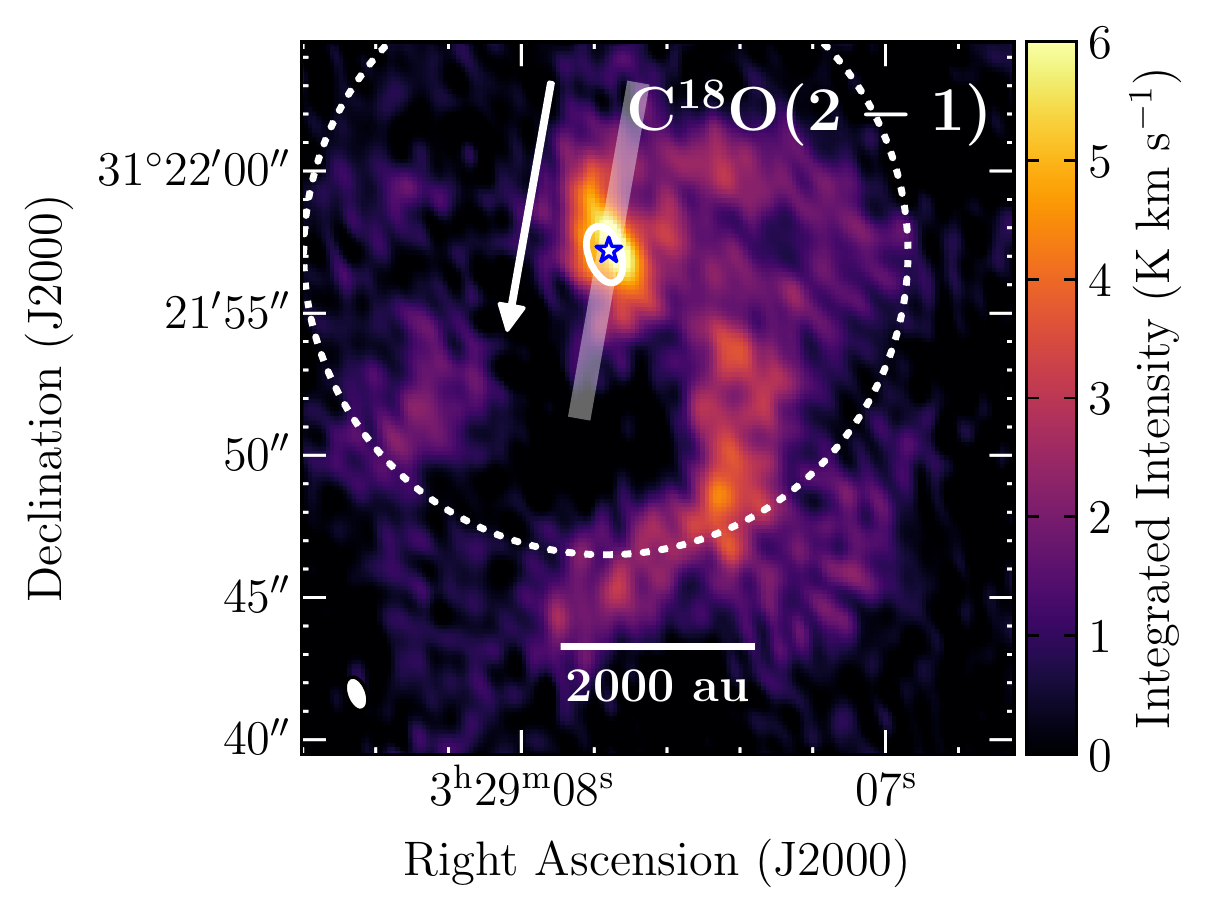}
     \includegraphics[width=0.48\textwidth]{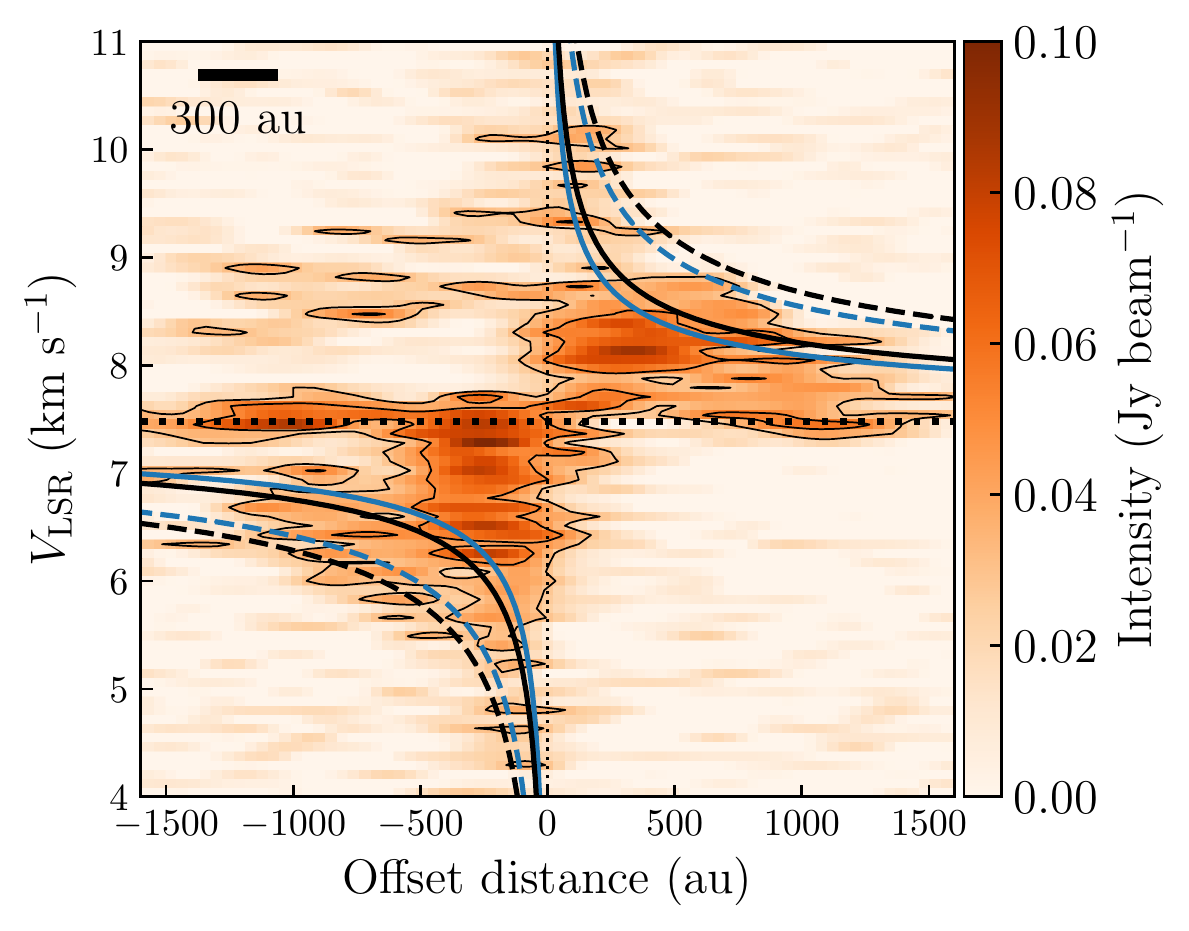}
    \caption{\label{fig:keplerC18O} C$^{18}$O integrated emission before primary beam correction alongside its PV diagram. \textbf{Left:} Integrated intensity image of C$^{18}$O between -1 and 14 \kms. The white semi-transparent line represents the position and width of the PV cut, built with the position angle of the disk from \cite{Segura-Cox2016}. The white arrow represents the direction of positive offset of the PV cut. The primary beam FWHM size is represented with a white dashed circle. The beam size is represented in a white ellipse in the bottom left corner. 
    \textbf{Right:} PV diagram of the C$^{18}$O line emission along the indicated PV cut. The horizontal dotted line represents Per-emb-50's $V_{LSR} = 7.5$ \kms. The vertical dotted line marks the central position of Per-emb-50. The PV diagram has an rms of 0.01 \Jyb. Intensity contours are placed at 5, 15 and 25 times the rms. The blue and black curves represent the Keplerian rotation curve for the minimum and maximum stellar mass, respectively, from \cite{Fiorellino2021-mdot}: The solid lines represent masses calculated assuming the star is located at the birthline of \cite{Palla1993birthmodel} model (0.53 --0.70 \Msun), whereas dashed lines represent the masses calculated for a 1 Myr protostar (1.52 -- 1.90 \Msun). Velocities are weighted according to the inclination angle as $V=V_{kep}\sin(i)$. The preferred Keplerian curve is the average between the dashed curves, with a mass of 1.7 \Msun. The scalebar at the top left represents a distance of 300 au, equivalent to the resolution of the NOEMA data.
    }
\end{figure*}

The integrated intensity image of C$^{18}$O\,(2 -- 1) is shown in Fig.~\ref{fig:keplerC18O} Left. We refer to C$^{18}$O\,(2 -- 1) as C$^{18}$O from now on, unless otherwise stated. The C$^{18}$O observations show the most extended emission of our NOEMA observations and have a similar velocity range as SO, between -1 and 14 \kms. This molecule's emission closest to the protostar allows us to determine the protostar's velocity and mass. 

We produce the position-velocity (PV) diagram for C$^{18}$O along the major axis of the disk in Per-emb-50 found by \cite{Segura-Cox2016} (170$^{\circ}$ counter-clockwise from North, see Fig.~\ref{fig:keplerC18O}). We use the astropy package \texttt{pvextractor} \citep{pvextractor} to obtain the PV diagram along a path centered on the protostar spanning a total length of 2400 au. We averaged all emission along the path that is within 1\arcsec width. The resulting PV diagram in Fig.~\ref{fig:keplerC18O} Right is consistent with rotation, with increasing velocity toward the protostar. The C$^{18}$O emission might be tracing Keplerian rotation.

Our observations of C$^{18}$O allow us to constrain the mass of the protostar. We obtain a central protostellar velocity $V_{\mathrm{LSR}}$ of 7.5 \kms, and a central protostellar mass $M_{*}=1.7\pm 0.2$ \Msun from the C$^{18}$O PV diagram. For this, we first manually determine the velocity that minimizes the asymmetries in the PV diagram. This results in a $V_{\mathrm{LSR}}=7.5$ \kms, marked with a horizontal dotted line in Fig.~\ref{fig:keplerC18O}. Afterwards, we compare the PV diagram with the Keplerian rotation curves produced by the masses previously estimated for Per-emb-50 by \cite{Fiorellino2021-mdot} using IR spectroscopy: they obtained a range between $0.5-0.7$ \Msun for a star located at the birthline at the HR diagram (using \cite{Palla1993birthmodel} model) and $1.5-1.9$ \Msun for a 1 Myr old protostar. The Keplerian rotation curves are weighted according to the inclination angle as $\mathrm{v}=\mathrm{v}_{\mathrm{kep}}\sin(i)$, where $i=67^{\circ}$ \citep[see Table \ref{tab:peremb50},][]{Segura-Cox2016}, with $i=0^{\circ}$ corresponding to a face-on disk.
The Keplerian rotation curves for a central protostar of 1\,Myr present a good correlation with the 3$\sigma$ contours ($\sigma=14$ m\Jyb, see Table \ref{tab:cubeprops}) of the C$^{18}$O PV diagram. We use the average between the 1 Myr mass upper and lower limits, 1.7 \Msun, and their difference as uncertainty ($\pm0.2$ \Msun).



\subsection{Streamline model\label{sec:streaminemodel}}

We model the kinematics of the streamer observed with H$_2$CO emission to confirm that the velocity gradient observed in H$_2$CO emission is consistent with infall motion, using the analytic solution for material falling from a rotating, finite-sized cloud toward a central object, dominated by the gravitational force of the latter. We use the analytic solutions of \cite{Mendoza2009}, previously used by \cite{Pineda2020} on the Per-emb-2 streamer. The model returns the position $\Vec{x_i}=(x,y,z)_i$ in au and velocity $\Vec{V_i}=(\mathrm{v}_x,\mathrm{v}_y,\mathrm{v}_z)_i$ in \kms (in Cartesian coordinates) of a parcel of mass along a streamline, where the z axis is defined along the angular momentum vector of the disk and the x-y plane is the disk plane. The model's input is the initial position and radial velocity of the parcel of mass within the cloud in spherical coordinates (initial radial distance $r_0$, position angle $\vartheta_0$ with respect to the z axis, inclination angle $\varphi_0$ which marks the initial angle within the disk plane and radial velocity $\mathrm{v}_{r,0}$) and the initial angular velocity of the cloud $\Omega_0$. We also apply two rotations due to the inclination angle $i$ and position angle $PA$ of the disk, to obtain the position and velocity with respect to the observer's point of view from the disk's reference system. 

The streamline model requires as input the central mass that dominates the gravitation of the system, which in this case is the sum of the masses of the protostar, disk and envelope, $M_{tot}=M_{*}+M_{env}+M_{disk}$. We use $M_{*}=1.7\pm0.2$ \Msun (see Sect. \ref{sec:protostellarmass}) and $M_{disk} = 0.58$ \Msun, the upper limit calculated in \cite{Segura-Cox2016}. For the envelope mass we use an upper limit of 0.39 \Msun and a lower limit of 0.18 \Msun, obtained using the Bolocam 1.1 mm image from \cite{Enoch2006}, taking the emission of Per-emb-50 with the disk component removed (see Appendix \ref{ap:envelopemass} for details).

We manually input the initial position ($r_0$, $\vartheta_0$ and $\varphi_0$), velocity $\mathrm{v}_{r,0}$ and inclinations $i$ and $PA$ to find the best parameters. We first assume that the streamer's rotation direction, given by $i$ and $PA$, are the same as the 
dust disk $i$ and $PA$ from \cite{Segura-Cox2016} (see Table \ref{tab:peremb50}). The inclination angle $i$ obtained from the dust disk is degenerate in 3-dimensional space (it can be inclined in $67^{\circ}$ or $-67^{\circ}$). We use the rotation direction given by the C$^{18}$O PV diagram in Sect. \ref{sec:protostellarmass} and the outflow direction (see Fig.~\ref{fig:images} Left) to determine that the angular velocity vector of the disk $\Vec{\omega}$ points toward the west (in the direction of the blueshifted outflow component) and is inclined toward the observer, thus $i=-67^{\circ}$. Then, we attempt to find analytic solutions with other $i$ and $PA$ values. The $i$ and $PA$ from \cite{Segura-Cox2016} and our disambiguation give the only rotation direction where we could find a solution for the velocity profile of the streamer.

\begin{figure*}[htbp]
    \centering
    \includegraphics[width=0.45\textwidth]{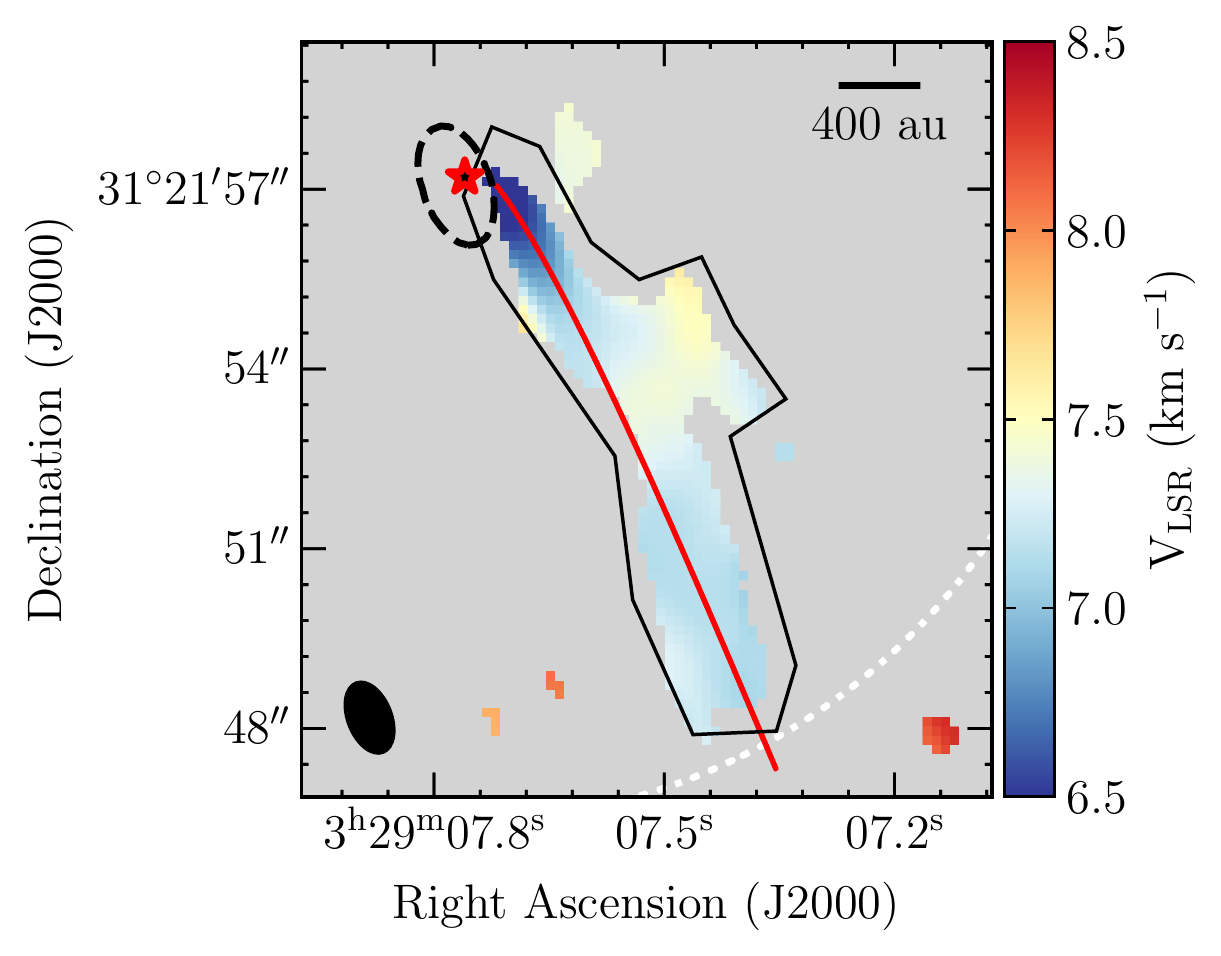}
    \includegraphics[width=0.45\textwidth]{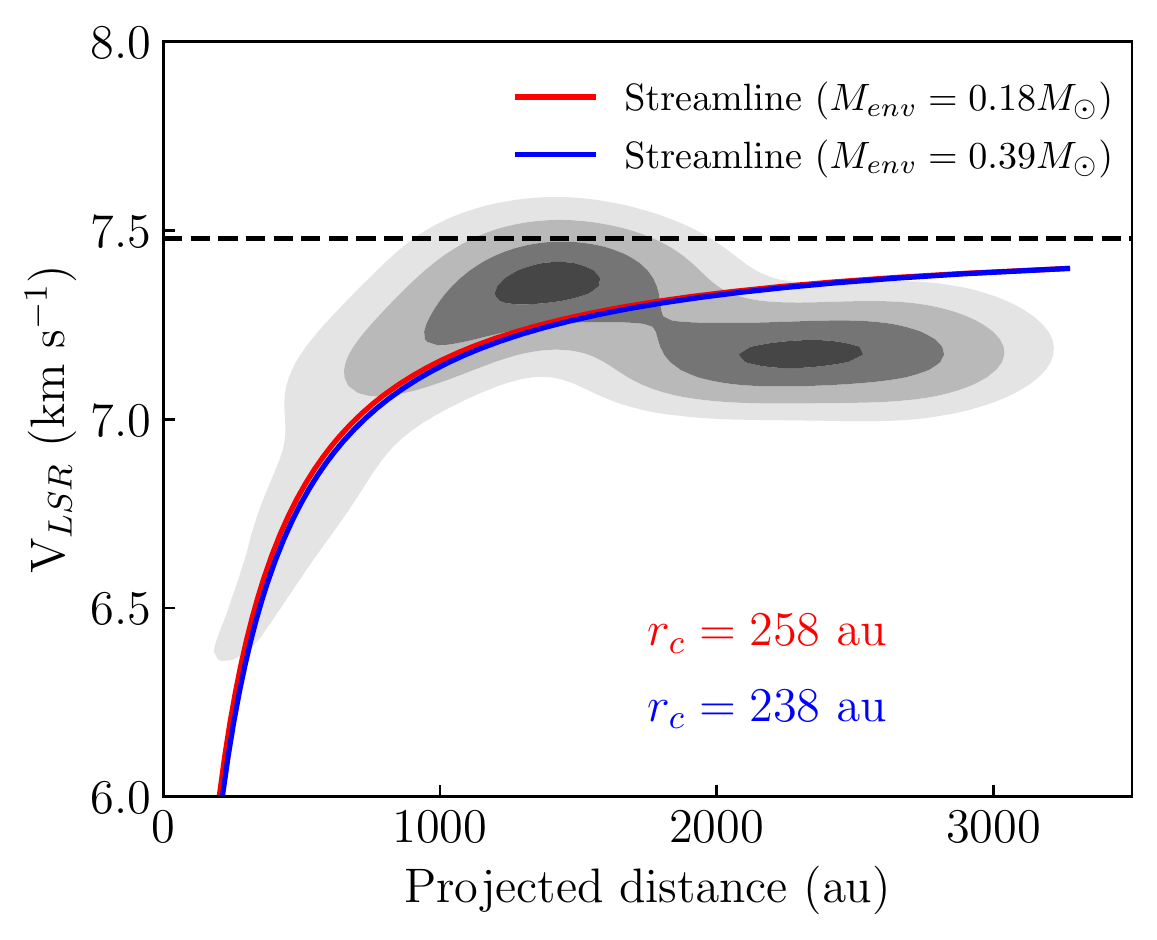}
    \caption{Central velocity of the best fit for each spectra in the H$_2$CO line emission, together with the streamline model for H$_2$CO. The red and blue lines represent the best streamline model for an envelope mass of 0.18 \Msun (total central mass of 2.47 \Msun) and 0.39 \Msun (total mass of 2.68 \Msun), respectively.\textbf{ Left:} central velocities with respect to position in the sky. The black polygon represents the region where velocities are sampled to build the kernel density estimation (KDE). The primary beam FWHM size is represented with a white dashed circle. The beam size is represented in a black ellipse in the bottom left corner. \textbf{Right:} Central velocity KDE as a function of projected distance to Per-emb-50 is plotted in grayscale. The KDE levels are drawn from 0.5$\sigma$ in steps of 0.5$\sigma$, where the $\sigma$ level is equivalent to the standard deviation of a normalized bivariate normal distribution. The dashed horizontal line represents the central velocity of Per-emb-50 $V_{\mathrm{LSR}} = 7.5$ \kms. The centrifugal radius resulting from each envelope mass are located below the curve.}
    \label{fig:streamerH2CO}
\end{figure*}

Table \ref{tab:paramsstream} lists the parameters that result in the analytic solutions for an infalling mass that best reproduce the H$_2$CO line profiles in the image plane and the structure of the velocity along the line of sight. 
Figure \ref{fig:streamerH2CO} shows the projected trajectory of the streamline model with the best parameters over the central velocity of the Gaussian fit to the H$_2$CO emission, both in the image plane (Left panel) and over the kernel density estimate (KDE) of the velocity and projected distance in the data (Right panel). We use the KDE implementation in the python package \texttt{SciPy} \citep{2020SciPy-NMeth} over the resulting central velocities obtained in Sect. \ref{sec:H2COgausfit}. 
The streamline model is able to reproduce the general shape of the KDE and the acceleration toward the protostar in the inner 1000 au. The model is not able to reproduce the slight discontinuity seen in the KDE at $\sim1700$ au, which is related to the kink feature (see Sec. \ref{sec:H2COgausfit}). The difference between using the upper and lower limits of the envelope mass is negligible in both the image and velocity planes (red and blue curves in Fig.\ref{fig:streamerH2CO}). 

\begin{table}[ht]
\centering
\caption{\label{tab:paramsstream}Parameters of the streamline model that reproduce best the H$_2$CO observations.}

\begin{tabular}{lll}
\hline\hline
Parameter  & Unit   & Value     \\ \hline
$\vartheta_0$ & deg   & 61.5  \\
$\varphi_0$   & deg    & 28.0  \\
$r_0$      & au     & \commentstere{3330}   \\
$v_{r,0}$  & \kms   & 1.25   \\
$\Omega_0$ & s$^{-1}$ & $4.53 \times 10^{-13}$ \\
$i$        & deg   & -67    \\
P.A.       & deg  & 170  \\ \hline
\end{tabular}
\end{table}

The centrifugal radius $r_c$ \citep[called $r_u$ in ][]{Mendoza2009} given by the parameters in Table \ref{tab:paramsstream} is between $r_c=238-258$ au, using the upper or lower limit for the envelope mass, respectively, both of which are within the beam size. This radius is the limit where the streamer can be modeled as free-falling matter with constant angular momentum, so we interpret this radius as approximately where the streamer deposits its material. This implies that the streamer deposits its mass at a distance about 150 au from the gas disk's edge, which we estimate has a radius of approximately 90 au using the SO line emission obtained in this work (see Sect. \ref{sec:SOdecomposition}). We do not use the streamline model solutions for distances smaller than $r_c$, as the model does not include motions within the gas and dust disk.

\subsection{Streamer mass\label{sec:streamermass}}

We calculate the streamer's mass and infall rate using the primary beam corrected C$^{18}$O emission in the area where the streamer is detected in H$_2$CO emission, as we can convert C$^{18}$O emission to gas emission with simple assumptions. \commentstere{We use the primary beam corrected emission because we use the intensity of the C$^{18}$O line, which we obtain by multiplying the map by the primary beam response, whereas in Sect. \ref{sec:H2COgausfit} we only need the central velocity and velocity dispersion of each spectrum to characterize the streamer's kinematics.}

C$^{18}$O emission is the most extended of all the molecular transitions used in this work, as it traces not only the gas in the streamer, but also the extended gas in the inner envelope and the filament where the protostar is \commentstere{embedded into}, which has a larger extension than the FoV. Nevertheless, the streamer is easily detected in C$^{18}$O, with a $\mathrm{S/N}\approx 10$ at the streamer's tail.
The C$^{18}$O emission shows a similar structure as in the H$_2$CO map. We cannot characterize the C$^{18}$O extended emission and kinematics outside of the streamer as we lack zero-spacing data, and we see some negative bowls in the image (see the black areas in Fig.~\ref{fig:keplerC18O} Right), indicating missing flux from larger scales. Therefore, for this work we use C$^{18}$O emission to describe the protostar and streamer's mass only.  

\begin{figure}[ht]
     \centering
     \includegraphics[width=0.45\textwidth]{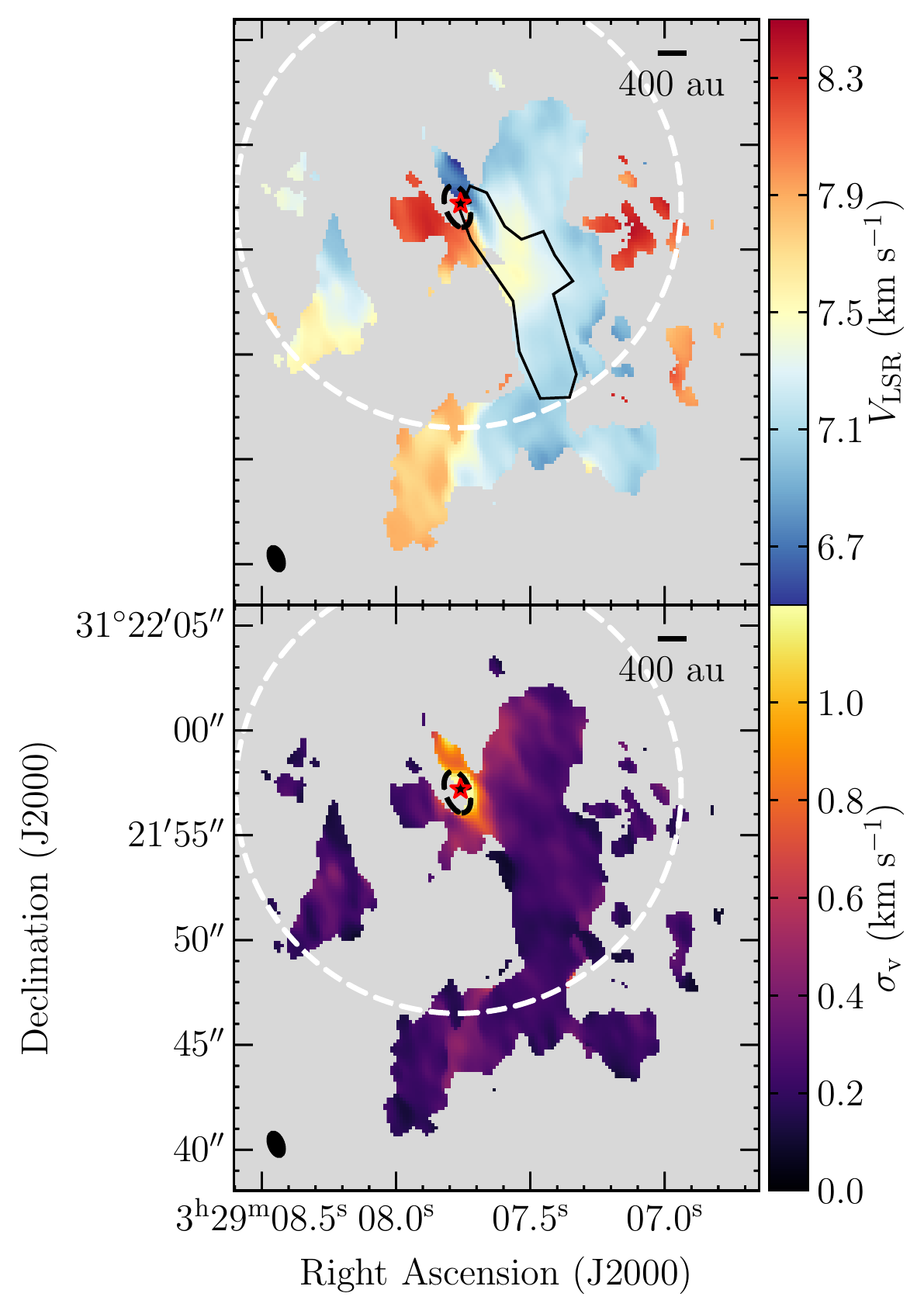}
     \caption{Velocity  $V_{\mathrm{LSR}}$ and velocity dispersion $\sigma_{\mathrm{v}}$ of the best Gaussian fit for C$^{18}$O emission. Gaussian fits are only done for spectra with $\mathrm{S/N}>4$. The primary beam FWHM size is represented with a white dashed circle. The beam size is represented in a white ellipse in the bottom left corner. \textbf{Top:} Central velocity of the best fit Gaussian profile for each spectrum. The black contour shows the same region plotted in Fig.\,\ref{fig:H2COfit} Right, where the mass is calculated from. \textbf{Bottom:} Velocity dispersion of the best fit Gaussian profile for each spectra.}
        \label{fig:C18Ofit}
\end{figure}

C$^{18}$O shows a similar central velocity as H$_2$CO at the streamer's tail but different velocity distribution at the position of the protostar, as shown in Fig.~\ref{fig:C18Ofit}. We use the same procedure for H$_2$CO to obtain the best Gaussian that fits the spectrum of each pixel with $\mathrm{S/N}>4$, described in Appendix \ref{ap:gaussfit}. Where the emission is coincident with H$_2$CO, C$^{18}$O is well described with one Gaussian component which shares the same $V_{\mathrm{LSR}}$ and $\sigma_{\mathrm{v}}$ as H$_2$CO emission (compare Fig.~\ref{fig:H2COfit} with Fig.~\ref{fig:C18Ofit}). The kink in velocities observed in the middle of the streamer is also observed in C$^{18}$O. Surrounding the protostar, outside of the area traced by the continuum, we observe blueshifted emission toward the northwest and redshifted toward the east. This emission probably traces a mixture of part of the inner envelope and disk rotation and the inner section of the outflow, as it follows the same east-west direction as the $^{12}$CO outflow detected by \cite{Stephens2019}. Therefore, it is safe to use C$^{18}$O emission within the region used to characterize the streamer's kinematics (black polygon in Fig.~\ref{fig:H2COfit} and Fig.~\ref{fig:C18Ofit}) to determine the streamer's mass. 

We obtain a mass lower limit for the streamer within the region drawn in Figures  \ref{fig:H2COfit} and \ref{fig:C18Ofit}. We detail the reasons why this is a lower limit in Sect. \ref{sec:discussion-streamlowlims}. We calculate the mass within the streamer assuming that C$^{18}$O is optically thin, under local thermodynamic equilibrium (LTE), and the streamer has a constant temperature $T_{ex}$. We use the values in the vicinity of Per-emb-50 in \cite{Friesen2017} and \cite{Dhabal2019}, which are between 10 and 20 K, hence we assume $T_{ex}=15\pm 5$ K.
First, we obtain the column density of the C$^{18}$O molecule, $N(\mathrm{C}^{18}\mathrm{O})$, using the primary beam corrected C$^{18}$O image. We explain the details of this procedure in Appendix \ref{ap:columndens}. The C$^{18}$O column density is around $2.8\times10^{15}$ cm$^{-2}$ within 1000 au of the protostar, then it falls to $\approx 8.0\times10^{14}$ cm$^{-2}$ and in the outer 1500 au it reaches up to $3.6\times10^{15}$ cm$^{-2}$. Afterwards, we transform $N(\mathrm{C}^{18}\mathrm{O})$ to molecular Hydrogen column density $N(\mathrm{H}_2)$ using $N(\mathrm{H}_2) = X_{\mathrm{C}^{18}\mathrm{O}} N(\mathrm{C}^{18}\mathrm{O})$. We use the canonical ratio $X_{\mathrm{C}^{18}\mathrm{O}}=5.9 \times 10^{6}$ \citep{Frerking1982}. 
Finally, we obtain the gas mass in the streamer using:
\begin{equation}
    M_{\mathrm{streamer}} = M_{gas} = \mu m_{H} d^2\, \delta x\, \delta y \sum N_{\mathrm{H}_2}, \label{eq:mh2mass}
\end{equation}
where $\sum N_{\mathrm{H}_2}$ is the sum of $N_{\mathrm{H}_2}$ in the streamer in cm$^{-2}$, $d$ is the distance to the protostar in cm, $\delta x\, \delta y$ is the size of the pixels in radians, $\mu=2.7$ is the molecular weight of the gas, considering the contribution from H$_2$, He and heavy elements, and $m_{H}$ is the H atom mass. We use $d = 293 \pm 22$ pc, the distance to NGC 1333 \citep[][see Table \ref{tab:peremb50}]{Ortiz-Leon2018}. 

We obtain a lower limit for the total mass of the streamer $M_{\mathrm{streamer}} = 1.2 \times 10^{-2}$ M$_{\odot}$, with an uncertainty of 15\% due to uncertainties in flux calibration and in the distance to NGC 1333 (see Table \ref{tab:peremb50}). 

\subsection{Streamer infall rate\label{sec:streamerinfallrate}}

We calculate the mean infall rate and the infall rate along the streamer using the mass obtained in Sect. \ref{sec:streamermass}, and compare it to the protostellar accretion rate. We differentiate between infall rate $\Dot{M}_{in}$, which is the rate at which mass is deposited from the envelope to the disk scales, and accretion rate $\Dot{M}_{acc}$, which is the rate at which the protostar is accreting mass.

\begin{figure}
\centering
\includegraphics[width=0.49\textwidth]{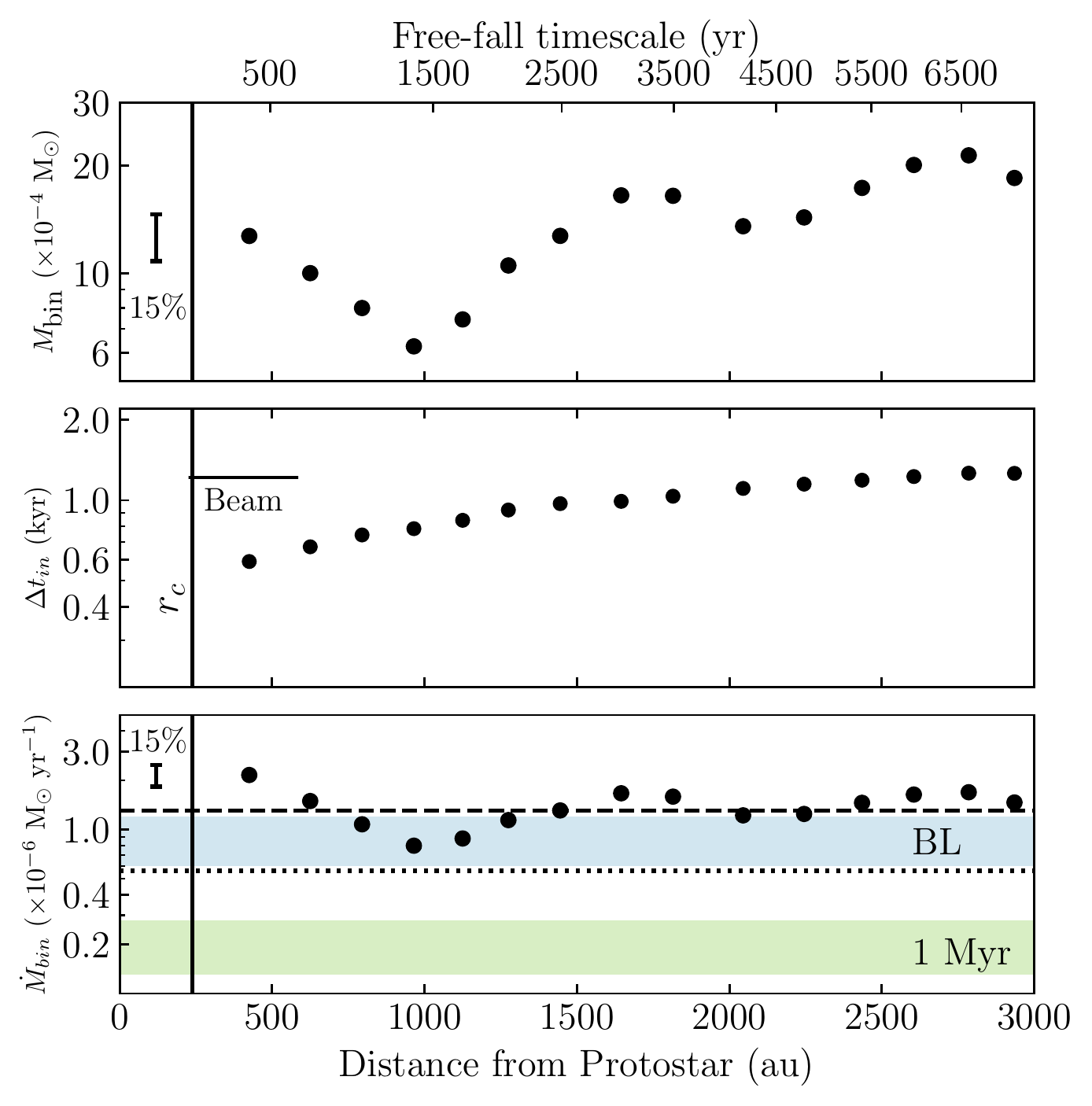}
    \caption{Mass, timescale and infall rate with respect to the distance to the protostar along the streamline. $\Delta t_{in}$ is obtained from the model with $M_{env}=0.39$ \Msun. The dotted line represents the mean infall rate obtained with the free fall timescale from rest, $\langle \Dot{M}_{in}\rangle = 5.6\times10^{-7}$ \Msun yr$^{-1}$, whereas the dashed line marks the mean infall rate obtained from the streamline model, $\langle \Dot{M}_{in, \mathrm{model}}\rangle = 1.3\times 10^{-6}$ \Msun yr$^{-1}$ (see Table \ref{tab:streamermass}). The blue and green areas correspond to the estimated accretion rates at the birthline (BL) and after 1 Myr of birth \citep{Fiorellino2021-mdot}. Systematic errors are plotted as vertical bars and are 15\% of the nominal values. Random errors represent around 5\%.
    \label{fig:massaccretion}}
\end{figure}

The free-fall timescale of the streamer, assuming the classic free-fall time equation,
\begin{equation}
    t_{\mathrm{ff}} = \sqrt{\frac{R^3}{GM_{tot}}},\label{eq:freefall}
\end{equation}
is $21.3\pm0.8$ kyr for an envelope mass of 0.18 \Msun ($M_{tot}=2.47$ \Msun), and $20.5\pm0.7$ kyr for $M_{env}=0.39$ \Msun ($M_{tot}=2.68$ \Msun). In Equation \ref{eq:freefall}, $M_{tot}$ is the total mass within a distance $R=r_0=3300$ au from the protostar (obtained from the streamline model in Sect. \ref{sec:streaminemodel}), and $G$ is the gravitational constant. We divide the total mass with the free-fall timescale to obtain an average $\langle \Dot{M}_{in}\rangle$ between $(5.4-5.6)\times10^{-7}$ \Msun yr$^{-1}$. The upper limit is plotted as a dotted line in Fig.~\ref{fig:massaccretion}.

Since we constrain the streamer's kinematics (see Sect. \ref{sec:streaminemodel}) and its column density at each position, we now derive the infall rate at every position of the streamer.
We first calculate the free-fall timescale $t_{\mathrm{ff}, \mathrm{model}}$ and average infall rate $\langle \Dot{M}_{in, \mathrm{model}}\rangle$ using the analytic solutions from Sect. \ref{sec:streaminemodel} to compare it to the classical free-fall timescale $\langle \Dot{M}_{in} \rangle$. For this, we calculate the travel time along the streamer by using the streamer's trajectory and velocities from the streamline model, from $r_0=3300$ au to the centrifugal radius, which we assume is the landing point (we use $r_c=238$ au). 
We obtain a total free-fall time of 8.7 kyr for $M_{env}=0.18$ \Msun and 8.6 kyr for $M_{env}=0.39$ \Msun, around 2 times lower than the times calculated previously, because the classic free-fall timescale (Equation \ref{eq:freefall}) does not consider that the streamline already has an initial velocity toward the protostar at $R$. The resulting infall rate is $\langle \Dot{M}_{in, \mathrm{model}}\rangle = 1.3\times 10^{-6}$ \Msun yr$^{-1}$ for both envelope masses. The average $\langle \Dot{M}_{in, \mathrm{model}}\rangle$ using the streamline model is plotted as a dashed line in Fig.~\ref{fig:massaccretion}. The mass and average infall rates found for the streamer are summarized in Table \ref{tab:streamermass}.

\begin{table}
\caption{\label{tab:streamermass}Global properties of the streamer found in Sect. \ref{sec:streamermass} and \ref{sec:streamerinfallrate}.}
\centering
\begin{tabular}{lll}
\hline\hline
Property  & Unit   & Value     \\ \hline
$M_{\mathrm{streamer}}$ & \Msun   & $(1.2\pm0.2) \times 10^{-2}$  \\
$t_{\mathrm{ff}}$ & kyr   & 20.5 -- 21.3  \\
$t_{\mathrm{ff},\mathrm{model}}$ & kyr   & 8.6 -- 8.7  \\
$\langle \Dot{M}_{in}\rangle$   & \Msun yr$^{-1}$    & $(5.4-5.6)\times10^{-7}$  \\
$\langle \Dot{M}_{in, \mathrm{model}}\rangle$    & \Msun yr$^{-1}$      & $1.3\times 10^{-6}$   \\ \hline
\end{tabular}
\end{table}

We also study how the infall rate changes along the streamer, to determine if there are significant differences in the infall rate within the streamer. Figure \ref{fig:C18Ofit} Left shows that molecular emission is clumpy on scales of the beam size, which suggests that there might be small-scale variations along the streamer. We separate the streamer into radial bins and obtain the mean 3-dimensional distance to the protostar $r_{bin}$, the total mass $M_{bin}$, the time taken to traverse the bin $\Delta t_{bin}$ and the infall rate $\Dot{M}_{bin}$ in each bin. The bins are 360 au wide (which is the major axis FWHM of the beam) and consist of all pixels that are within a certain range of projected distances $[r, r+360]$ au from Per-emb-50.  We sample every 120 au (1/3 of the major axis of the beam) from 200 au to 3300 au from the protostar, in projected distance. The resulting mass, crossing time and infall rates for each bin are in Fig.~\ref{fig:massaccretion}.

We calculate $r_{bin}$ as the distance of the streamline model point that is closest to the center of mass of the bin in the image plane. We use $N(\mathrm{C}^{18}\mathrm{O})$ to find the center of mass within each bin and then find the point in the streamline model closest to it. Then, the distance $r_{bin}$ is the 3-dimensional distance between that point and the protostar. We express this distance as the free-fall timescale from $r_{bin}$ using:
\begin{equation}
    t_{bin} = -\int_{r_{bin}}^{0} \frac{dr'}{\sqrt{\mathrm{v}_{r,0}^2 + 2GM_{tot} \Big(\frac{1}{r'} - \frac{1}{r_0}\Big)}}, \label{eq:integralffwithv0}
\end{equation}
where $\mathrm{v}_{r,0}$ is the initial velocity (1.25 \kms) at $r_0$ (3300 au) from the streamline model toward the direction of the protostar. The integral is done numerically using the python package \texttt{scipy} function \texttt{integrate}. The difference between the solution of Equation \ref{eq:integralffwithv0} and the free-fall timescale given by the streamline model is less than 20 yr, which is negligible for the timescales we are working with. 

We compute the infall rate of the streamer using the mass within each bin $M_{bin}$ and the time it takes to cross the bin $\Delta t_{in}$. $M_{bin}$ is calculated using Equation \ref{eq:mh2mass}, adding $N_{\mathrm{H}_2}$ in all pixels that belong to the bin. We then calculate  $\Delta t_{bin}$ the same way as the total free-fall timescale, but adding up the time obtained from the trajectory and velocities within the bin only. The infall rate for each bin is therefore $\Dot{M}_{bin} = M_{bin}/\Delta t_{bin}$. 

The infall rate along the streamer is consistently larger or equal to the accretion rate estimated for Per-emb-50, independent of the variations along the streamer. Figure \ref{fig:massaccretion} shows the resulting $M_{bin}$, $\Delta t_{bin}$ and $\Dot{M}_{bin}$ with respect to the distance to the protostar $r_{bin}$, and compares the infall rates $\Dot{M}_{bin}$ with the accretion rates $\Dot{M}_{acc}$ for Per-emb-50 estimated in \cite{Fiorellino2021-mdot}. The average $\Dot{M}_{bin}$, $\langle \Dot{M}_{in}\rangle$, estimated using the streamline model is between 5-10 times larger than the $\Dot{M}_{acc}$ estimated for a 1 Myr protostar ($(1.3-2.8)\times10^{-7}$ \Msun yr$^{-1}$), and just above the upper limit for the $\Dot{M}_{acc}$ of Per-emb-50 assuming it is located at the Birth-line of the \cite{Palla1993birthmodel} model ($1.2\times10^{-6}$ \Msun yr$^{-1}$). The protostellar mass calculated in Sect. \ref{sec:protostellarmass} is consistent with a 1\,Myr protostar, so likely the accretion rate is the former, resulting in  $\langle \Dot{M}_{in}\rangle / \Dot{M}_{acc}=5-10$. Therefore, the streamer is feeding more than enough mass to sustain the accretion rate of the protostar, and according to our total free-fall time, we can expect a similar infall rate for at least the next 8.7 kyr.

The mass per bin varies from $6\times10^{-4}$ to $2\times10^{-3}$ \Msun from bin to bin. This variation drives the fluctuations observed in the infall rates, which are within a factor of 3, \commentstere{with minima located at $\sim$1000 and $\sim$2000 au}. Nevertheless, these variations are small and the streamer shows a consistently high infall rate along its full length, reflected in $\langle \Dot{M}_{in}\rangle$. The fluctuations are present in spatial scales larger than 300 au, so these are not affected by the resolution limit. \commentstere{The mass variations might be because the streamer is clumpy, with changes in scales smaller than our 300 au resolution. On the other hand, the MRS of the data is around 22\arcsec, but the data are already less sensitive to extended emission before reaching that distance, at around 4\arcsec, so the apparent minima in the infall curve of Fig.~\ref{fig:massaccretion} might be explained by a decreased sensitivity to extended sources. }

\subsection{Asymmetries in SO and SO$_2$ emission\label{sec:SOpvdiag}}

Figure \ref{fig:images} Right shows the SO integrated emission obtained with NOEMA. Unlike H$_2$CO and C$^{18}$O, SO emission in Per-emb-50 is brightest at about 150 au south of the protostar, 
and extended out to around 1000 au from it (see also Fig.~\ref{fig:SOspectra}). 
The southern part of the SO emission overlaps with the brightest H$_2$CO emission. It also presents emission at $\gtrapprox$ 3000 au from the protostar, but since this emission lies outside the primary beam, we will not describe it further in this work.  SO is known to be a tracer of cold, dense gas \citep[e.g.,][]{Swade1989L134n_chem, HacarTafalla2011Taurus} and it is sublimated from dust grains by sufficient heating, for example, by accretion shocks around the centrifugal barrier \citep[e.g.,][]{Sakai2014Nature, vanGelder2021SO}.
SO is found in young, embedded sources, but not in T Tauri disks \citep{Guilloteau2016ChemOfDisks, Dutrey2011SInTTauri}, suggesting an increasing S depletion with disk age.
This hints that SO is tracing the dense inner envelope and gas disk around the protostar.  

We use SO$_2$\,(11$_{1,11}$ -- 10$_{0,10}$) emission to aid in the interpretation of the SO emission. The SO$_2$ integrated intensity image is shown in Fig.~\ref{fig:SO2withspectra}, together with selected spectra. SO$_2$ emission is compact and peaks at the south of Per-emb-50, 
close to where H$_2$CO emission ends. Its peak is at the same location as the SO peak, but its emission is $\sim5$ times weaker than SO. The SO$_2$ molecule is a known shock tracer as it can trace warm areas in accretion shocks \citep{vanGelder2021SO}, in particular at the disk-envelope surface \citep[e.g., ][]{ArturDeLaVillarmois2019ClassIOph}. This suggests that the SO$_2$ emission in the south of Per-emb-50 might be tracing shocked material, probably due to either the streamer impacting zone or another phenomena. 

\begin{figure}[htbp]
     \centering
    \includegraphics[width=0.5\textwidth]{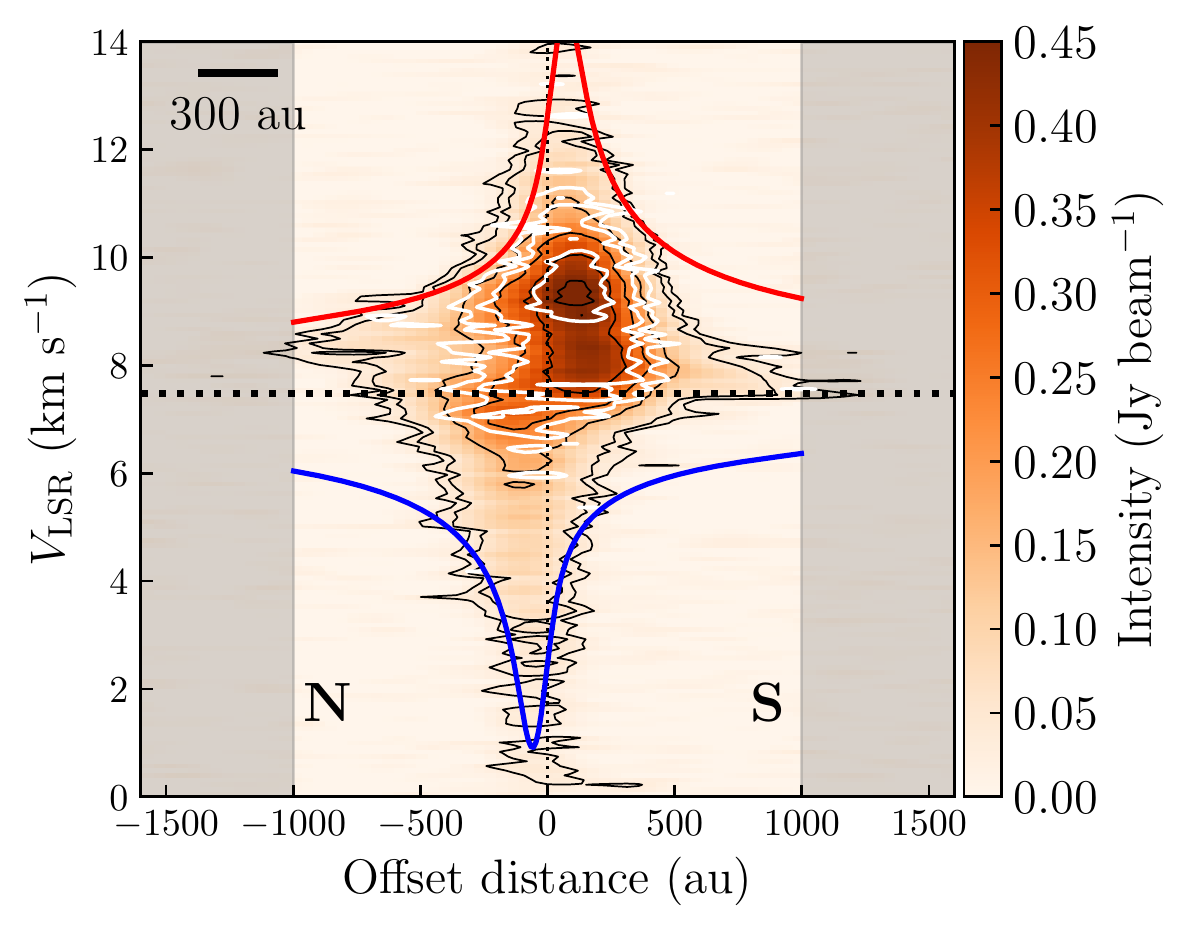}
    \caption{\label{fig:pvSO-SO2} Position-Velocity diagram of SO(5$_5$ -- 4$_4$) (color and black contours) and SO$_2$($11_{1,11}-10_{0,10}$) (white contours) line emission along the position angle of the disk from \cite{Segura-Cox2016}, plotted along the same offset scale as Fig.~\ref{fig:keplerC18O} Right. Negative offsets represent distances toward the North and positive offsets toward the South. The horizontal dotted line represents Per-emb-50's V$_{LSR} = 7.48$ \kms. The vertical dotted line marks the central position of Per-emb-50. Both PV diagrams have an rms of 0.01 \Jyb. Intensity contours for SO are placed at 3, 5, 15, 25, 35 and 45 times the rms, whereas SO$_2$ intensity contours are placed at 3, 5 and 10 times the rms. The red and blue solid curves show the model of rotation and infall that best matches the SO for the redshifted and blueshifted velocities with respect to the protostar, respectively. The scalebar at the top left represents a distance of 300 au, equivalent to the resolution of the NOEMA data.
    }
\end{figure}

We generate the PV diagrams of SO and SO$_2$ line emission along the same cut done for C$^{18}$O in Sect. \ref{sec:protostellarmass} to investigate what kinematics are these molecular lines tracing. The resulting PV diagrams are shown in Fig.~\ref{fig:pvSO-SO2}. The shapes of both PV diagrams differ from the C$^{18}$O PV diagram (see Fig.~\ref{fig:C18Ofit}), indicating that these molecules trace different kinematic components. SO has a skewed diamond-shaped emission,  with both blueshifted and redshifted components at the north and south parts of the cut, which suggests mixture of infall and rotation motions, whereas C$^{18}$O has a bowtie shape consistent with motion dominated almost entirely by Keplerian rotation. Additionally, the brightest SO emission comes from redshifted velocities both toward the north and south of Per-emb-50, whereas blueshifted emission comes almost fully from the northern side of the inner envelope. Unlike SO, SO$_2$ emission is only present around the peak, with no recognizable characteristic shape and barely presents emission over $3\sigma$ at blueshifted velocities. Both diagrams peak at the same position, within the inner 300 au from the protostar toward the southeast, and in velocity, at approximately 9.5 \kms. The shape of these two molecules' emission suggest they follow motions which are asymmetric in the north-south direction. 

We fit the ``toy model" for infall and rotation motion from \cite{Sakai2014Nature} to the SO PV diagram to investigate if the diamond shape is consistent with the rotation and infall kinematics of a flattened inner envelope. The free parameters in this model are the centrifugal radius of the material in the envelope $r_{c, \mathrm{env}}$ (not to be confused with the centrifugal radius of the streamer $r_c$) and the mass of the central object $M_{tot}$. The best fit curves from this toy model are plotted in red and blue for the redshifted and blueshifted sides, respectively, overlaid on top of the SO PV diagram in Fig.~\ref{fig:pvSO-SO2}. The model must be divided in two parts to be able to reproduce the shape of the diagram: the redshifted and blueshifted side are best fitted with a different set of parameters. The redshifted side is best fitted with a toy model with $M_{tot, \mathrm{r}}=4$ \Msun and $r_{c, \mathrm{env}, \mathrm{r}}=130$ au, whereas for the blueshifted side $M_{tot, \mathrm{b}}=2.9$ \Msun and $r_{c, \mathrm{env}, \mathrm{b}}=100$ au. Therefore, SO molecular emission is tracing asymmetric kinematics in the inner envelope consistent with infall and rotation, where the redshifted emission (which is brighter) possesses a different motion than the blueshifted side. The fact that the masses $M_{tot, \mathrm{r}}$ and $M_{tot, \mathrm{b}}$ are higher than the protostellar mass we determined kinematically (1.7 \Msun, see Sect. \ref{sec:protostellarmass}), plus the fact that they are different, suggests that the model does not capture all the kinematic phenomena in the envelope. \commentstere{These results lead us to investigate the SO emission in more detail.}




\subsection{Gaussian components of SO emission\label{sec:SOdecomposition}}

The complex shape of the SO PV diagram, the strong peak at redshifted velocities, and the fact that it can be fitted with the \cite{Sakai2014Nature} toy model with two different set of parameters for the redshifted and blueshifted parts, suggest that there are at least three components being traced: rotation, infall and a strong, redshifted component. We separate the different kinematic components through Gaussian spectral fitting of SO to study them separately. 

We fit one, two and three Gaussians to the SO spectrum of each pixel with $\mathrm{S/N}>4$ using the same method for H$_2$CO and C$^{18}$O emission, described in Appendix \ref{ap:gaussfit}. Figure \ref{fig:SOspectra} shows four spectra in different regions with their respective best fit curves. Most of the SO spectra require two Gaussians, or in some cases, three Gaussians to be reproduced. For each pixel, we evaluate  how much improvement we obtain by adding a second and third Gaussian using the Akaike Information Criterion (AIC, see Appendix \ref{ap:gaussfit} for details). With the decomposed spectra, we investigate the separate physical components of SO emission that can be described using each Gaussian.

\begin{figure*}[htbp]
     \centering
     \includegraphics[width=0.49\textwidth]{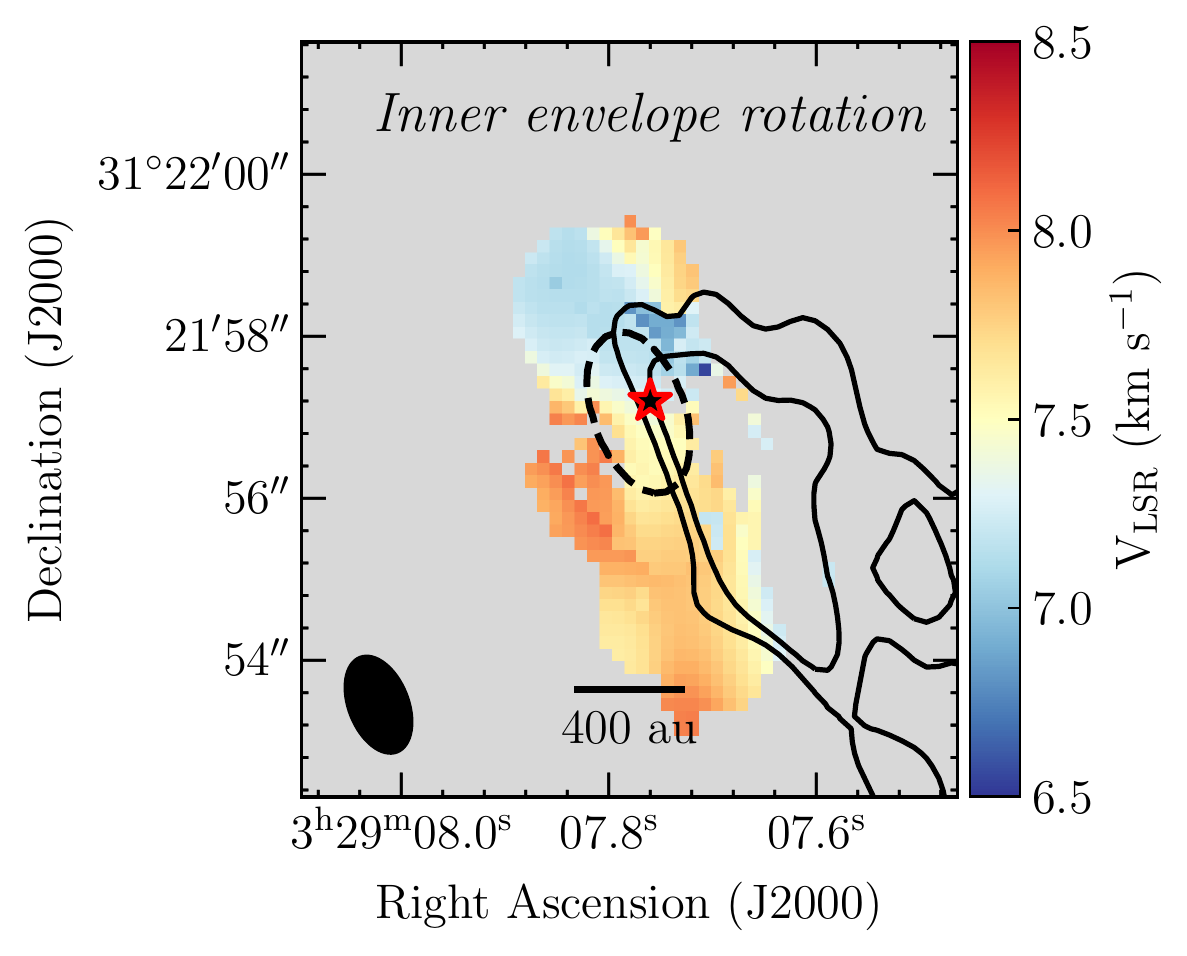}
     \includegraphics[width=0.49\textwidth]{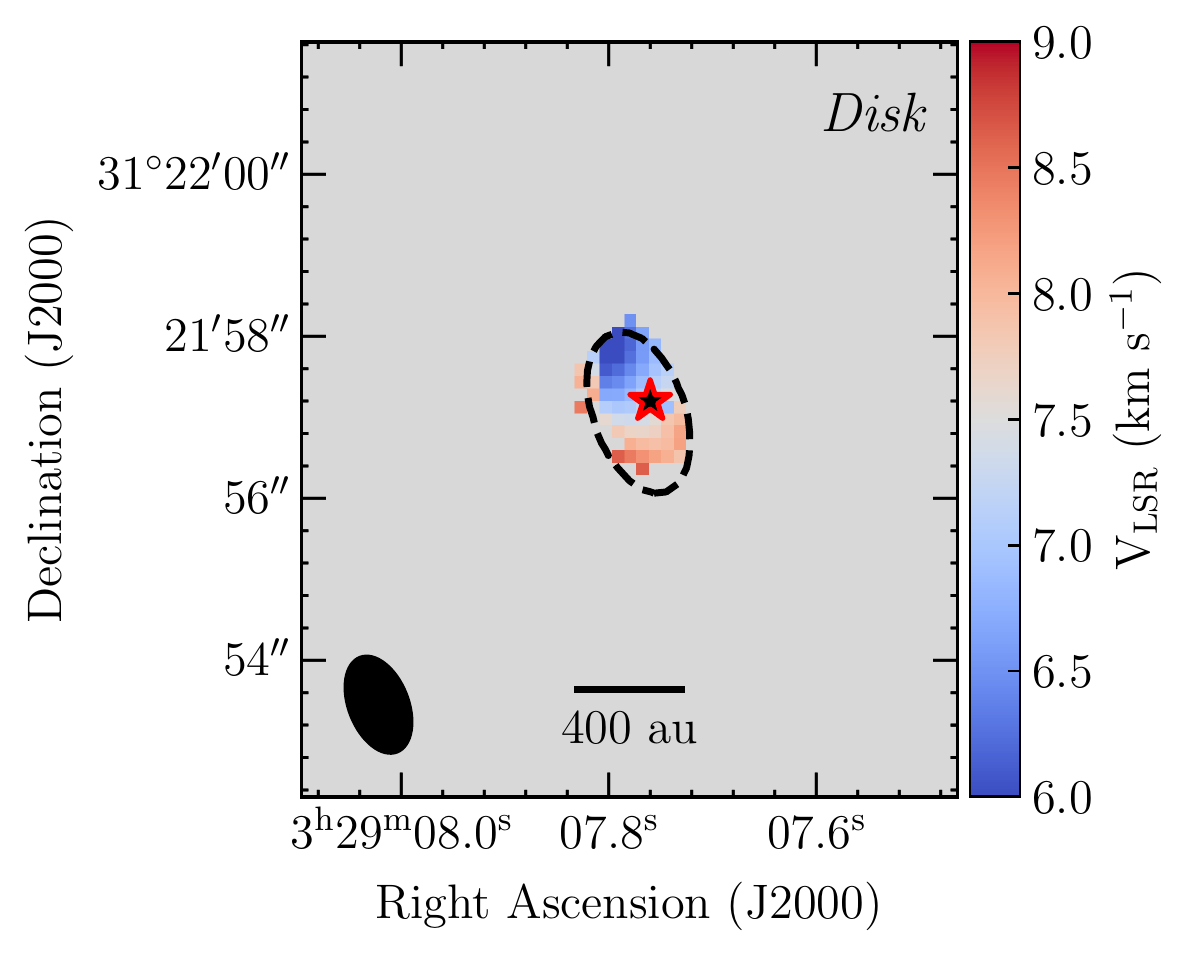}
     \includegraphics[width=0.49\textwidth]{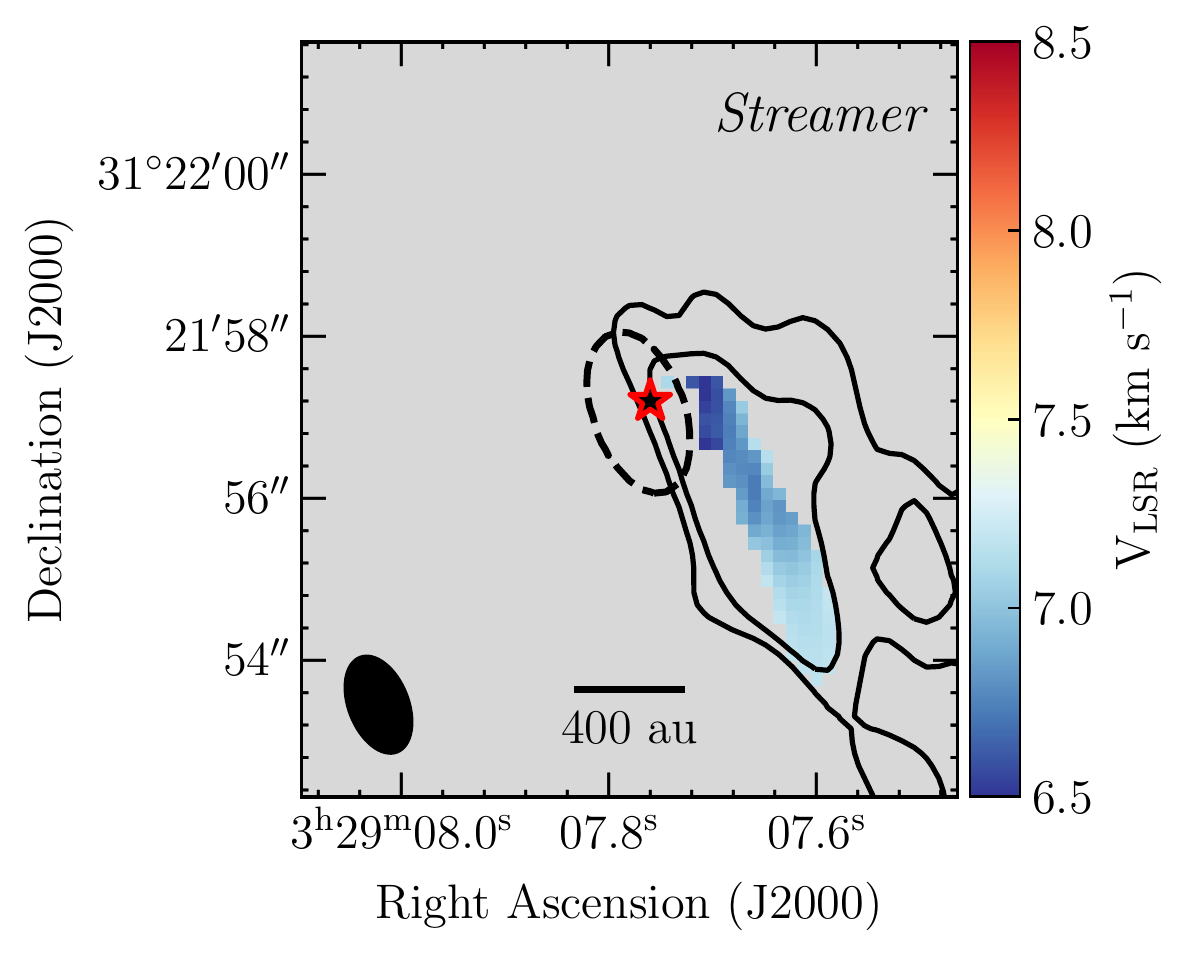}
     \includegraphics[width=0.49\textwidth]{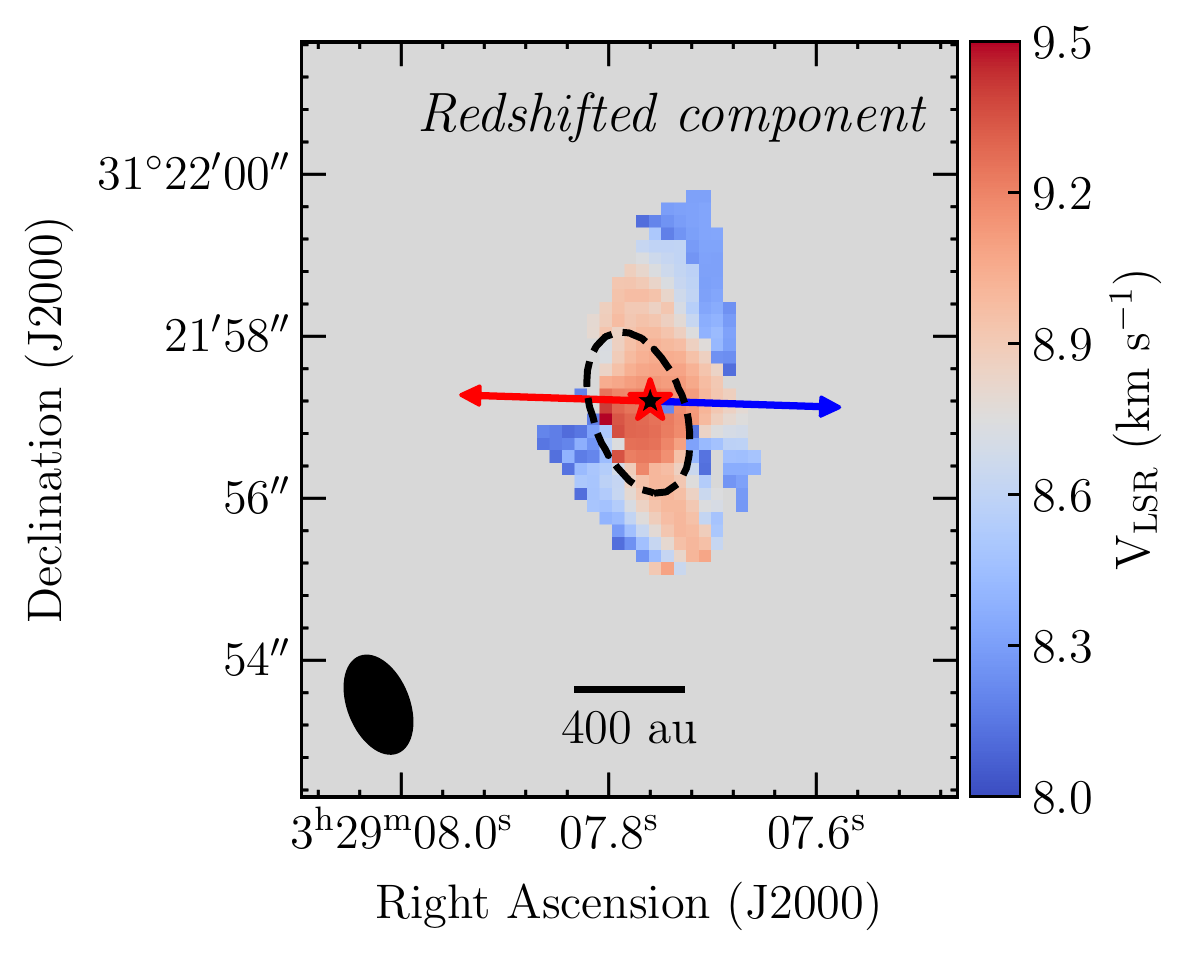}
    \caption{\label{fig:SOcomponents} Best fit central velocity maps for the 4 signature components found in SO (5$_5$-4$_4$) emission in the inner 1000 au of Per-emb-50's envelope. Examples of the individual spectra decomposition for each of these components are shown in Fig.~\ref{fig:SOspectra}. Black solid contours represent the integrated intensity of H$_2$CO emission at 3 and 5 times the rms of the integrated intensity map (0.25 K \kms). Black dashed contours mark the continuum emission at a 7 m\Jyb level (see Fig.\,\ref{fig:cont}). The red and blue arrows indicate the outflow direction determined by the redshifted and blueshifted CO (2 -- 1) emission, respectively, from \cite{Stephens2019}. The clean beam is represented as a black ellipse in the bottom left corner of all images. Note that the velocity ranges plotted in each figure are different, which is why the figures on the right have a different color scheme. \textbf{Top Left:} inner envelope rotation.\textbf{Top Right:} broad feature ($\sigma_{\mathrm{v}} > 2$ \kms). \textbf{Bottom Left}: streamer. \textbf{Bottom Right}: redshifted component.}
\end{figure*}

We find 4 signature components in SO emission: one consistent with inner envelope rotation, a compact feature around the protostar with a large velocity dispersion ($\sigma_{\mathrm{v}} > 2$ \kms), a third component consistent with the streamer's kinematics from Sect. \ref{sec:asym-inflows} and a fourth component completely redshifted with respect to Per-emb-50. We separate each of the components using the following steps:

\begin{enumerate}
    \item All Gaussian curves which have a velocity dispersion $\sigma_{\mathrm{v}} > 2$ \kms correspond to the broad feature, which is consistent with \textit{marginally-resolved disk rotation}.
    \item Then, all Gaussians with $\sigma_{\mathrm{v}} < 2$ \kms which have a central velocity $V_{LSR}> 8.1$ \kms correspond to the strong, \textit{redshifted component}.
    \item The Gaussians left which have $V_{LSR}< 7.2$ \kms and are located at a lower declination than $+31^{\circ}21'57.6"$ are consistent with the \textit{streamer}.
    \item All pixels left contain only one Gaussian curve, which has a central velocity map consistent with \textit{inner envelope rotation}.
\end{enumerate}

The central velocity $V_{\mathrm{LSR}}$ of the 4 separated components are shown in Fig.~\ref{fig:SOcomponents}. We show the best fit parameters (central velocity and dispersion) for each component in Appendix \ref{ap:SOdecomp-results}. Figure \ref{fig:SOspectra} shows the components in selected SO spectra. 



The inner envelope rotation component contributes to the diamond shape shown in the PV diagram in Fig.~\ref{fig:pvSO-SO2} with the blueshifted emission on the northern side and the redshifted emission in the southern side. This rotation component is resolved in our observations, extending a factor of $\sim 2$ farther in radius than the continuum emission (see Fig.~\ref{fig:SOcomponents} Top Left), so it does not correspond to disk rotation, and has the same rotation direction shown in our C$^{18}$O data (see Fig.~\ref{fig:keplerC18O}). 

Within the continuum emission contour, the SO spectra present emission fitted with Gaussians with blueshifted and redshifted velocities with respect to Per-emb-50 and with $\sigma_{\mathrm{v}} > 2$ \kms (see Fig.~\ref{fig:SOcomponents} Top Right and Fig.~\ref{fig:aux-SOsigmav} Top Right). The observed gradient in its central velocities is consistent with rotation kinematics, with the same rotation direction suggested by the C$^{18}$O PV diagram (see Fig.~\ref{fig:keplerC18O}) and the inner envelope rotation. However, as this component only emits within the continuum emission, we assume this gas belongs to the gas disk only, unlike C$^{18}$O which also traces the flattened inner envelope rotation. Using the stellar mass obtained in Sect. \ref{sec:protostellarmass} and the velocity dispersion from this SO component, we estimate the radius of this compact component assuming it traces Keplerian rotation and that at the disk edge the Keplerian velocity is $v_k\sim \sigma_{\mathrm{v}}\approx 4$ \kms. This estimate returns a disk radius of approximately 90 au. Therefore, this component is consistent with a gas disk around the protostar.  


Towards the south of Per-emb-50, one of the fitted Gaussian components is consistent with the streamer structure found in H$_2$CO, both in position in the sky and velocity (compare Fig.~\ref{fig:H2COfit} and Fig.~\ref{fig:SOcomponents} Bottom Left). This component is clearly separated from all other components in the south as it is blueshifted with respect to the protostar's  $V_{\mathrm{LSR}}$, whereas the other component close-by (inner envelope rotation) is redshifted (see Fig.\,\ref{fig:SOcomponents} and Fig.\,\ref{fig:SOspectra} Left).  This component's SO spectra shows the same central velocity as H$_2$CO (see Spectra d in Fig.~\ref{fig:SO2withspectra}) and acceleration toward blueshifted velocities found in the H$_2$CO Gaussian fitting (see Fig.~\ref{fig:SOcomponents} Bottom Left). SO traces only the inner 1000 au of the streamer, likely tracing its denser regions. 





The fourth component found through Gaussian decomposition is strongly redshifted with respect to the protostar (see Fig.~\ref{fig:SOcomponents} Bottom Right). This component has a larger velocity close to the center of the continuum emission (around 9.5 \kms) and decreases radially (to approximately 8.0 \kms). Its radial velocity gradient is not consistent with the direction of the outflow nor the streamer. We propose this component might be tracing another asymmetric infall, located along the line of sight. This infall is asymmetric as we do not see a strongly blueshifted counterpart ($V_{LSR}< 7$ \kms) covering a similar area, expected for an axisymmetric infall. The only strongly blueshifted component is very thin and located in the same area as the streamer. With the present observations, we do not have enough spatial resolution to characterize this infall further.




\section{Discussion\label{sec:discussion}}

\subsection{Why are mass and infall rate lower limits?\label{sec:discussion-streamlowlims}}

The estimated mass of the streamer (see Sect. \ref{sec:streamermass}) is a lower limit because of observational limits in our data and the assumptions made in the mass calculation. 

We estimate the length of the streamer as 3300 au, using H$_2$CO emission and the streamline model. This is possibly not the full length of the streamer for three reasons. First, the H$_2$CO emission is cut off by the primary beam of the NOEMA observations ($22\arcsec$), and our observations are not sensitive to strong emission beyond this radius. Second, there is a strong offset emission toward the southwest of Per-emb-50, located just outside the primary beam at $\sim 3000$ au, seen in all of the molecular tracers used in this work (see Figs. \ref{fig:images} and \ref{fig:C18Ofit}). Moreover, there is significant C$^{18}$O emission observed in the SMA MASSES program \citep{Stephens2019} in the same location as the H$_2$CO streamer, which extends to a bright emission located beyond the streamer's observed extent in this work.  Third, the streamline model requires an initial velocity $\mathrm{v}_{r,0}=1.25$ \kms in the direction of the protostar to fit the outer 1500 au of the streamer (see Table \ref{tab:paramsstream}). The initial velocity might indicate that the streamer starts farther away and was already infalling by the time it reaches $r_0$. 

Another observational limitation is the lack of zero-spacing data. C$^{18}$O emission is extended and the observations have no sensitivity to scales larger than the MRS (22\arcsec), but our observations start losing sensitivity to scales larger than 4\arcsec due to the coverage in u-v space. Therefore, we are not certain if the clumpiness observed in C$^{18}$O is real or it is influenced by missing flux due to lack of zero-spacings. 

The main assumptions that we use in the streamer's mass calculation are, first, a fixed ratio between column densities which is suitable for an undepleted gas, $X_{\mathrm{C}^{18}\mathrm{O}} = 5.9\times10^{6}$ \citep{Frerking1982}, and second, we assume a constant excitation temperature $T_{ex}$. Most likely, $X_{\mathrm{C}^{18}\mathrm{O}}$ is not constant along the streamer. 
Within the dense core, it is more probable that there is a larger C$^{18}$O depletion into grains due to an increase in density \citep[see ][and references within]{BerginTafalla2007coldcloudsreview}. Where C$^{18}$O is depleted, $X_{\mathrm{C}^{18}\mathrm{O}}$ should be higher to estimate the mass correctly. Also, this conversion factor is calibrated using Taurus molecular clouds, and might differ in Perseus. \cite{Pineda2008PerseusCO} show that there is variation in the conversion factors of the C$^{18}$O(1--0) line in different regions in Perseus. Secondly, a constant $T_{ex}$ along the streamline is unlikely: the temperature might be higher closer to the protostar due to thermal heating. This is also suggested by the presence of SO$_2$ emission toward the south of Per-emb-50. Unfortunately, we do not have a good estimation of the gas temperature in the vicinity of Per-emb-50. NH$_3$ is a commonly used chemical thermometer, combining the (1,1) and (2,2) inversion transitions, both observed in NGC 1333 with GBT \citep{Friesen2017}. Although the NH$_3$(1,1) line is present in Per-emb-50, the NH$_3$(2,2) line is too faint to be detected around the protostar and provide a gas temperature estimation. Higher spatial resolution observations of both NH$_3$ lines do not detect emission in this region \citep{Dhabal2019}. Instead, we use the values in the vicinity of Per-emb-50 in \cite{Friesen2017} and \cite{Dhabal2019}, which are between 10 and 20 K. The variance in $T_{ex}$ adds less than 5\% of the total uncertainty, and therefore it does not dominate the uncertainties. 

Given that the mass and mass infall rates we report are lower limits, the general results of this paper are strengthened: the streamer delivers more than enough mass toward the protostellar disk to sustain its high accretion rate in comparison with its neighbors (see Sect. \ref{sec:streamermass} and Sect. \ref{sec:discussion-outbursts}). If the streamer masses or infall rates are actually higher, this streamer can deliver even more mass than what we report here.

\subsection{Classical free-fall time versus streamline model\label{sec:discussion-fftime}}

For the first time, we calculate the infall timescales along a streamer using the streamline model based on the analytical solution from \cite{Mendoza2009}. We show that in Per-emb-50, where the streamline model requires an initial radial velocity, the infall rate is underestimated by at least a factor of 2 when calculated with the classic -- and initially static -- free-fall timescale. The factor by which the timescale is underestimated depends on the initial velocity of the streamer: if the streamer presents an initial impulse at the starting radius $r_0$, it will take less time to reach the protostellar disk than if the streamer started from rest. The streamline model allows to estimate the initial radial velocity. This highlights the importance of the use of a streamline model to calculate the timescales involved in infall.

The calculation of the initial radial velocity (and consequently, the infall rate) relies on a streamer model that has good constraints both spatially in the image plane and kinematically in the velocity along the line of sight. If the streamer is fully contained along the line of sight, the velocity is correctly characterized but we do not have information about the length of the streamer. On the other hand, if the streamer moves completely within the plane of the sky, there is information about the length and path of the streamer, but the velocity cannot be characterized. Fortunately, in the case of Per-emb-50, the streamer is mostly contained in the plane of the sky, with a small inclination at the start of the streamline (approximatelt 10$^{\circ}$ according to the streamline model in Sect. \ref{sec:streaminemodel}), and it becomes more inclined with respect to the line of sight where we see the acceleration closer to the disk. This allows us to sample both the distance (up to the primary beam edge) and the velocity, and therefore constrain the initial radial velocity.





\subsection{Streamer is landing within disk scales \label{sec:discussion-impactzone}}

Our results indicate mass is infalling to disk scales (which corresponds to distances of $\sim 100-200$ au), both in the case of the streamer and the asymmetric infall seen in the redshifted component of SO emission (see Sect. \ref{sec:SOdecomposition}).
We can model the streamer down to $\approx 250$ au from the edge of the gas disk (see Sect. \ref{sec:streaminemodel}) and the toy model in Sect. \ref{sec:SOpvdiag} has a centrifugal radius between 100 and 130 au, similar to the 90 au of the gas disk. 
It is possible that SO$_2$ traces the impact zone where gas is infalling, either that of the streamer or the redshifted SO component. H$_2$CO and SO emission tracing the streamer end within a beam size of the location of the SO$_2$ peak emission, located at $\sim 150$ au (see Fig.\,\ref{fig:SO2withspectra}), which is compatible with the centrifugal radius obtained for the streamline model is $\approx 250$ au (see Sect. \ref{sec:streaminemodel}) as the emission is seen in projected distance and $r_c$ is a 3-dimensional distance. 
According to the streamline model, the impact velocity component along the line of sight at the assumed impact location ($r_c$) is $1.7$ \kms. The FWHM of the SO$_2$ emission spectra at the location of the streamer's end is similar to the estimated impact velocity, suggesting that the impact of the streamer is responsible for the SO$_2$ velocity dispersion. However, SO$_2$ peaks at the same velocity as the strong, redshifted component that could be attributed to another asymmetric infall, and at the peak location, both have the same shape (see Fig.\,\ref{fig:SO2withspectra} Top Right). Therefore, it is unclear which infalling feature most influences the SO$_2$ emission. 


One interesting result is that the centrifugal radius of the streamer $r_c$ ($\sim 250$ au, see Sect. \ref{sec:streaminemodel}) is about twice the centrifugal radii obtained for the rotating-infalling envelope, $r_{c, \mathrm{env}, \mathrm{r}}=130$ au and $r_{c, \mathrm{env}, \mathrm{b}}=100$ au (see Sect. \ref{sec:SOpvdiag}). 
This suggests that the streamer and envelope have different origins and then the streamer might come from outside the dense core. The streamer component seen in the SO emission might indicate the entrance of the streamer to the inner envelope, where the latter is flattened and has a rotating and infalling motion of its own (represented by the redshifted component in Sect. \ref{sec:SOdecomposition}). For the streamer material to reach the centrifugal radius of the inner envelope, which is slightly larger than the gas disk radius (90 au, see Sect. \ref{sec:SOdecomposition}), and for its material to reach the gas disk, it must lose angular momentum, for example, through magnetic braking \citep{Mestel-Spitzer1956MagBraking, Mouschovias1980MagBraking, Basu-Mouschovias1994MagBraking}. Loss of angular momentum of material coming from $>10\,000$ au has been observed for Class 0 sources by \cite{Pineda2019SpAngMomCores} down to $\sim 1000$ au, becoming low enough to generate a rotationally supported disk in scales $<100$ au. 
Future high resolution observations can clarify the interaction between the streamer and the inner envelope for Class I sources.


\subsection{Relation between streamers and accretion outbursts \label{sec:discussion-outbursts}}

\commentstere{The presence of streamers with a high infall rate, like the one found toward Per-emb-50, are linked to accretion variability and luminosity outbursts. Simulations of turbulent molecular clouds suggest that infall from larger scales regulates the accretion toward the protostar, even in later phases than Class 0 \citep{Padoan2014Infallsim, Kuffmeier2018EpisodicAcc}. 
In the case presented in this work, the relation between the streamer and a (current or future) accretion burst is supported by the high accretion rate and luminosity of Per-emb-50 in comparison with other Class I protostars, as well as other asymmetric structures found toward current (and past) outbursting sources. } 

\commentstere{The streamer feeding Per-emb-50 might explain the high accretion rate 
of this protostar in comparison to other Class I sources in NGC 1333.  
Its $\Dot{M}_{acc}$ is $\sim$10$\times$ higher than for other Class I sources in 
NGC 1333 \citep{Fiorellino2021-mdot}, and the infall rate provided by the streamer 
is 5--10$\times$ larger than $\Dot{M}_{acc}$ (see Sect. \ref{sec:streamerinfallrate}), 
more than enough to replenish the mass consumed by accretion. 
The luminosity \citep[between 10 and 25 \Lsun,][]{Enoch2009, Dunham2015gouldbeltcatalog} and accretion rate are consistent with those of Class Is 
undergoing an accretion burst \citep{Hsieh2019-ALMA_Outbursts}. 
However, Per-emb-50's envelope mass is similar to those around other 
Class I objects \citep[at 2.2 \Msun,][]{Enoch2009, Agurto-Gangas2019}, 
and the streamer might be the key ingredient to sustain 
Per-emb-50’s high accretion rate. }

\commentstere{It is also possible that we are seeing the protostar in 
the onset of an accretion burst, as it is significantly brighter than 
other Class I protostars, or the streamer might produce one in the future. 
Since the streamer's infall rate is 5--10$\times$ larger than the current accretion rate, it is possible that in the future 9000 yr the accretion rate can grow up to one order of magnitude. 
This shows that streamers might provide a significant amount of mass for stellar accretion, and suggest that intense accretion events can take place during the Class I phase. Moreover, if more streamers in Class I protostars are found and their masses characterized \citep[e.g., in this work and ][]{Yen2014L1489Infall}, this also suggests that the main accretion phase of the protostar might extend beyond the Class 0 phase. }

\commentstere{Recent observations toward young stellar objects find a correlation between accretion bursts and infall from larger scales. Asymmetric structures of 1000 au length have been associated with some FU Ori protostars \citep{Liu2016FUOriandInfall}. Other protostars with a known accretion burst in the past, such as Per-emb-2 \citep{Pineda2020} and V883 Ori \citep{White2019V883Ori}, also have streamers with an infall rate higher than their accretion rate. For these sources, it is suggested that the large-scale infall regulates the episodic accretion. This might be the case for Per-emb-50 as well: we propose that the mass is delivered to the protostellar disk, which triggers a disk instability \citep[like a gravitational instability, as suggested by][]{Kuffmeier2018EpisodicAcc,Hennebelle2017SpiralsAccretion}, the mass is transported through the disk and afterwards is accreted by the protostar in a burst. 
This idea is supported by the disk’s mass in comparison to other disks: Per-emb-50’s dust disk has between 0.28--0.58 \Msun, around twice the mass seen in other Class I disks \citep{Segura-Cox2016, SeguraCox2018VANDAM}, which suggests that this disk might be accumulating mass coming from the streamer. 
Additionally, even if we are currently unable to resolve this disk, gravitational instabilities produced by infalling material have been suggested to account for the spiral structures found in the disks of other protostars, for instance, in IRAS16293--2422 B \citep[a Class 0 source,][]{Zamponi2021iras16293},  HH 111 VLA 1 \citep[a Class I source,][]{Lee2020InfallProdSpirals} and Elias 2-27 \citep[a Class II protostar,][]{Paneque-Carreno2021spiralsdisk}. 
Higher resolution observations of the gas disk around Per-emb-50 are required to study these possible instabilities.}

\subsection{Where does the streamer come from?\label{sec:discussion-streamerorigin}}

\begin{figure}[htbp]
    \centering
    \includegraphics[width=0.45\textwidth]{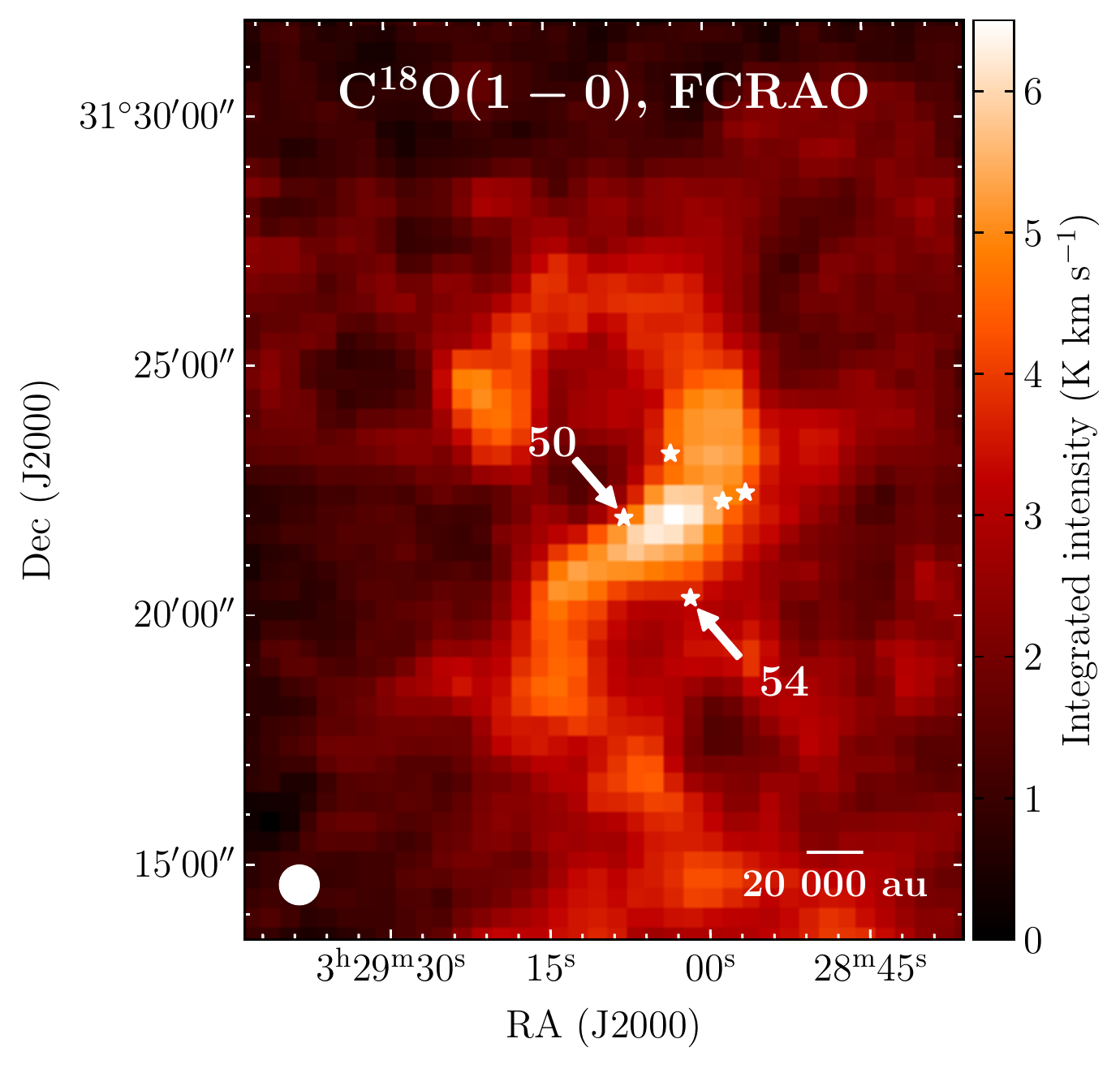}
    \caption{Integrated intensity map of C$^{18}$O(1--0) emission of NGC 1333 from \cite{Hatchell2005Perseus}, between 5 and 10 \kms, centered at the location of Per-emb-50. White stars represent the location of close-by Class I protostars. Per-emb-50 and Per-emb-54 are indicated and labeled with white arrows. The white circle in the lower left corner represents the beam of the C$^{18}$O(1--0) observations (46\arcsec).}
    \label{fig:C18Olargescale}
\end{figure}

The streamer possibly connects to larger scale structures such as filaments and fibers. Within molecular clouds, simulations suggest that that up to 50\% of the final protostellar mass comes from beyond their natal core \citep{Pelkonen2021massbeyondcore}, and observations of other protostars show that gas can flow from beyond the protostar's natal core, connecting the protostar with other structures \citep[e.g., ][]{Chou2016DiskandFilconnectionL1455}. Our data, together with the observed environment where Per-emb-50 lives, suggest that this might be the case for this protostar as well. 

First, as discussed previously (see Sect. \ref{sec:discussion-streamlowlims}), the H$_2$CO and C$^{18}$O emission are truncated by the NOEMA primary beam and there is significant C$^{18}$O (2 -- 1) emission observed in the SMA MASSES program \citep{Stephens2019}, located at the position of the offset emission outside the primary beam, directly in line with the streamer. Moreover, the MASSES emission is also cut short at its primary beam \citep[48\arcsec,][]{Stephens2019}.  
The gas reservoir seen in MASSES C$^{18}$O observations might be funneled by the streamer or be part of it, implying that streamers might connect with larger structures in their natal molecular clump.




Zooming out, NGC 1333 consists of a complex association of filaments, revealed in dense gas observations \citep{Chen2020fibers,Dhabal2019,Friesen2017}. At larger scales, the streamer points directly toward the crossing of two dense gas filaments observed in NH$_3$ observations \citep[filament b in][]{Chen2020fibers} and toward a bright extended emission source  
observed in C$^{18}$O (see Fig.~\ref{fig:C18Olargescale}), located between Per-emb-50 and Per-emb-54. If the streamer continues outside the primary beam, it may connect both protostars, as observed with the protostar L1544-IRS1 and the starless core HRF40 by \cite{Chou2016DiskandFilconnectionL1455}. There are currently no observations at intermediate resolution ($6-10\arcsec$) with an appropriate tracer in NGC 1333 that connects the large scale clumps and filaments surrounding Per-emb-50 to the core. Studies of filaments and fibers such as those of \cite{Chen2020fibers} and \cite{Dhabal2019} show an intricate connection between filaments and cores, but they are not sensitive enough to detect emission close to Per-emb-50, and the C$^{18}$O(1--0) has too coarse resolution \citep[46\arcsec beam,][]{Hatchell2005Perseus}. Nevertheless, the general direction of the streamer suggests that this streamer is connected to the larger scale filaments. 

\subsection{Asymmetries in SO and SO$_2$ emission}

The SO and SO$_2$ emission (see Fig.~\ref{fig:pvSO-SO2}) are asymmetrical: they are both brighter toward the south and in redshifted velocities. SO shows that the kinematics in this source are complex and include both asymmetric infall and rotation. This is more evident in the Gaussian decomposition (see Sect. \ref{sec:SOdecomposition}). These asymmetries show that the inner envelope of Per-emb-50 is not infalling monolithically, like the classical picture of core collapse \citep{Terebey1984rotation,Shu1977corecollapse}. 

Through Gaussian decomposition, we find that the redshifted component that dominates the SO emission is centered around the protostar and has a central velocity of approximately 9.5 \kms (2 \kms redshifted with respect to Per-emb-50, see Sect. \ref{sec:SOdecomposition}). We interpret this emission as another asymmetric infall completely contained within the line of sight. Given the velocity gradients seen in Fig.~\ref{fig:SOcomponents} Bottom Right, this component might not be a streamer but rather a wider asymmetric infall, comprising one side of the envelope located between the observer and the protostar. Finding a possible second infall feature in Per-emb-50 shows that the envelope infall kinematics are complex and reaffirms the idea that mass accretion does not follow an inside-out, axisymetric fashion. 
The asymmetries might be related to the environment where Per-emb-50 is located, close to the intersection of two filaments in NGC 1333 \citep{Chen2020fibers} and close to several other protostars \citep{Enoch2009, Dunham2015gouldbeltcatalog}.

\subsection{Comparison with other streamers\label{sec:discussion-comparison}}

\begin{figure*}[htbp]
    \centering
         \includegraphics[width=0.95\textwidth]{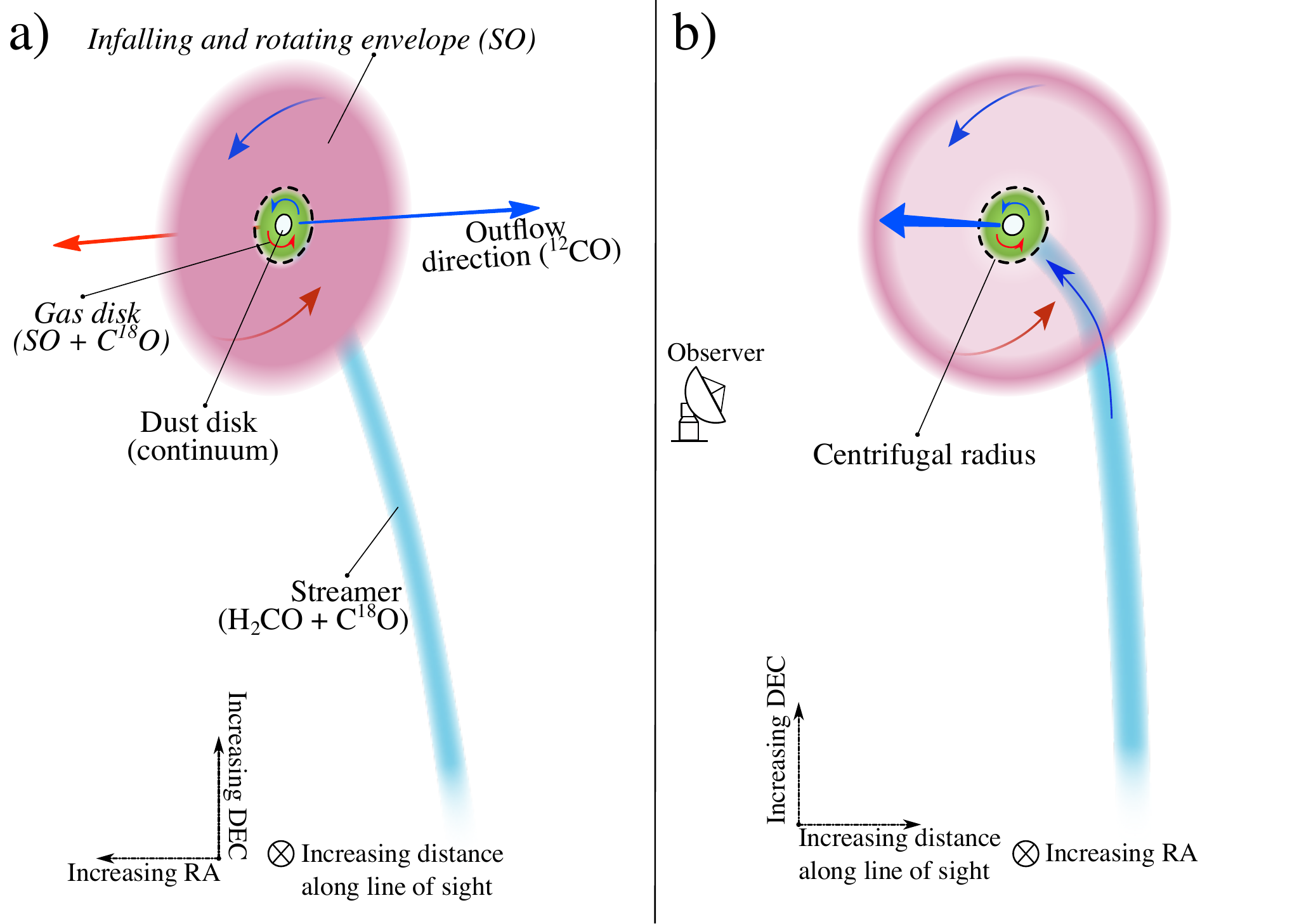}
        \caption{Schematic illustration of the different elements present around Per-emb-50. The streamer is represented as a light blue thick line. The inner envelope is represented in light pink and the rotating gas disk is represented with a green ellipse. The inner envelope has a transparent gap inside to see the streamer behind it. The blue and red straight arrows show the outflow direction in the blueshifted and redshifted directions, respectively. The white circle shows the dust disk and the solid black ellipse marks its edge. The dashed circle represents the centrifugal radius. Curved arrows show the general direction of motion of each component: red and blue arrows show if the motions are seen as redshifted or blueshifted line emission, respectively. \textbf{a:} Image plane view, where the right ascension increases toward the left and declination increases upwards. \textbf{b:} Line of sight view, where the observer is on the left and distance increases toward the right. }
        \label{fig:diagramPer50}
\end{figure*}

Streamers, defined as long ($\gtrsim 1000$ au) and asymmetric accretion flows toward disk-forming scales \citep[$\lesssim 300$ au, as in ][]{Pineda2020}, are a relatively new phenomenon which is proving relevant in star formation, with new discoveries both in gas tracers \citep[][Segura-Cox et al. in prep]{Alves2020} and dust \citep{Ginski2021}. Per-emb-50's streamer is the first Class I protostellar streamer to be characterized using a free-falling model. We illustrate the streamer and its relation with the various components found surrounding Per-emb-50 in Fig.~\ref{fig:diagramPer50}.

Per-emb-50's structure and kinematics are similar to other asymmetric features found in protostars in Perseus and other molecular clouds. The observed streamer size in this work is within the range of other observed streamers (between 1000 and 10\,000 au), such as toward [BHB2007]11 \citep{Alves2020}, Per-emb-2 \citep{Pineda2020} and 
SU Aur \citep{Ginski2021,Akiyama2019}. 
Similar infalling structures have been found at smaller scales (between roughly 200 to 1000 au), within inner envelopes of single systems \citep[e.g.,][]{Garufi2021arXiv-accretionDGTauHLTau,Tang2012ABAurLateAccretion} and within the circumbinary disk and inner envelope of binary systems \citep[e.g.,][]{Phuong2020GGTauSpirals,Takakuwa2017L1551NEBinarySpirals,Dutrey2014GGTau, Takakuwa2014L1551NEBinaryAngularMom}. The streamer in this work also shows a velocity gradient and a curved appearance in the image plane, as many of the streamers mentioned above \citep[e.g.,][]{Pineda2020,Akiyama2019}. We note that the infalling structures in smaller scales (200 -- 1000 au) might be of different nature, possibly driven by the tidal forces of the binary systems instead of pure free-fall. However, these structures also play a role in feeding the circumstellar disks. 

Our work uses the same analytical solution as in \cite{Pineda2020} for Per-emb-2, a Class 0 protostellar close binary ($<20$ au), the first streamer where mass and infall rate were obtained, but extends the method to include the analysis of infall rates along the streamer. In Per-emb-2, the streamer's kinematics are consistent with a model with $\mathrm{v}_{r,0}=0$, so using the free-fall timescale does not severely underestimate the infall rate. Per-emb-50's mean streamer infall rate $\langle \Dot{M}_{in}\rangle_{\mathrm{Per50}} = 1.3\times 10^{-6}$ \Msun yr$^{-1}$ is similar to the infall rate in Per-emb-2, $\Dot{M}_{in, \mathrm{Per2}}\approx 10^{-6}$ \Msun yr$^{-1}$ \citep{Pineda2020}. While the infall rate is similar in both sources, the mean ratio $\Dot{M}_{in}/\Dot{M}_{acc}$ is higher for Per-emb-50 (5-10, in contrast with 1.4 for Per-emb-2). Nevertheless, both are $>1$, even assuming the highest accretion rate possible for Per-emb-50, $(0.6-1.2)\times10^{-6}$ \Msun yr$^{-1}$ (blue area in Fig.~\ref{fig:massaccretion}).
Per-emb-50 is unique in that it is the first streamer to definitively show, through the use of a free-fall model, that the infall rate can sustain the accretion rate.
This implies that streamers can contribute important amounts of mass in later phases than Class 0, therefore suggesting that important accretion events can happen in the Class I phase, and in some cases, can occur in Class II sources \citep[as suggested by Garufi et al. 2021 subm., ][]{Alves2020, Tang2012ABAurLateAccretion}.

It is still uncertain if the lack of streamers found in observations is due to an observational bias, or streamers are uncommon in the majority of star forming systems. 
If streamers live as much as the estimated free-fall timescale of Per-emb-50 ($t_{ff}\sim9000$ yr), and the protostar has only one streamer in their life, there is a chance between 2\% and 30\% of observing one during the Class I phase: the lower limit obtained by dividing $t_{ff}$ by the estimated Class I phase duration \citep[0.44. Myr, ][]{Evans2009C2Dlifetime} and the upper limit dividing $t_{ff}$ by itself plus the time between accretion bursts, estimated to occur once every few 10 000 yrs \citep[][]{Frimann2017-MASSES_Outburst, Jorgensen2015accbursts}. This is just an order of magnitude estimate, as the time between bursts is uncertain and has a wide range of values in different protostars \citep[from few 1000 to few 10 000 yrs, e.g.,][]{Hsieh2018burstsinVELLOs, Frimann2017-MASSES_Outburst, Jorgensen2015accbursts} and previous works show this time might increase from Class 0 to Class I protostars \citep{Hsieh2019-ALMA_Outbursts, Audard2014OutburstsReview}.  Nevertheless, asymmetric infall features are seen along the complete simulations of star formation within a molecular cloud \citep{Kuznetsova2019, Kuffmeier2018EpisodicAcc}. 

As streamers are a new emerging phenomenon in observations, it is unclear which are the best molecules to trace them. Per-emb-50 shows the first streamer characterized with H$_2$CO emission, whereas other streamers are observed in $^{12}$CO \citep[e.g.,][]{Alves2020}, HC$_3$N \citep[e.g.,][]{Pineda2020} and HCO$^+$ \citep[e.g.,[]{Yen2019HLTau}. Upcoming NOEMA observations from the PRODIGE project can uncover more asymmetric infalls and streamers around Class 0/I sources and, in the future, we might be able to make a statistical study of streamers in protostars and investigate which molecules are the best tracers of this phenomenon.


\section{Conclusions\label{sec:conclusions}}

In this work, we present new NOEMA observations of H$_2$CO, C$^{18}$O, $^{12}$CO, SO and SO$_2$ molecular lines toward Per-emb-50, a Class I source in NGC 1333. We use these observations to characterize the kinematics from envelope to disk scales around the protostar. An illustration of our main findings is shown in Fig.~\ref{fig:diagramPer50}.
The main results are summarized as follows:

\begin{enumerate}
    \item We find a streamer depositing 
    material close to the edge of the gas disk around Per-emb-50. It presents a rougly constant velocity in H$_2$CO emission in the line of sight from roughly 1500 to 3000 au from the protostar. There is acceleration toward more blueshifted velocities closer to the protostar, up to around 1000 au. 
    
    \item The analytical solutions for infalling gas along a streamline can reproduce the observed kinematics of the H$_2$CO emission. An initial velocity of 1.25 \kms at the initial position \commentstere{3330} au away from the protostar is required to replicate the observed velocity along the line of sight. Taking the initial velocity into account, the free-fall timescale of the streamer is $\sim9000$ yr. 
    
    \item The streamer is delivering more than enough mass to sustain its protostellar accretion rate. We estimate a lower limit to the streamer's mass at $1.2\times10^{-2}$ \Msun, from which we obtain a mean infall rate of $1.3\times 10^{-6}$ \Msun yr$^{-1}$, with factor 3 variations along the streamer. The infall rate is consistently about 5 to 10 times larger than the estimated accretion rate of the protostar. This means that the streamer can deliver enough mass to sustain the high accretion rate of this protostar for at least the next 9000 yrs. 
    
    \item We find signatures of asymmetry in SO and SO$_2$ emission. The PV diagram of SO shows a diamond shape consistent with rotation and infall motions, but there is an asymmetry between the redshifted and blueshifted velocities. Through Gaussian decomposition, we find that SO is tracing mostly the inner envelope rotation and a redshifted asymmetric infall located along the line of sight. SO is also tracing the inner 1000 au of the streamer. SO$_2$ emission hints at an impact zone toward the south of Per-emb-50, which is consistent with both the estimated landing site of the streamer and the peak of the redshifted asymmetric infall. 
\end{enumerate}

The description of the envelope around Per-emb-50 and each of its distinct kinematic components 
are limited by the resolution and primary beam of our observations, together with the lack of zero-spacing data. We emphasize that the streamer might extend further than the 3000 au we characterize in this work, as it is traced in C$^{18}$O, which is observed outside of the primary beam of our observations, 
points toward the crossing of two dense gas filaments, and that the mass is a lower limit. Further observations with single dish antennas will allow to obtain the total flux (and therefore, mass) along the streamer and confirm its mass fluctuations. Intermediate resolution observations ($\approx6\arcsec$) that cover an area larger than the NOEMA primary beam will allow us to investigate the connection of this streamer to the larger filament. Higher spatial resolution data of more than one SO and SO$_2$ molecular transitions will help determine the precise landing site of the streamer and allow us to characterize the redshifted infall better. Observations of other transitions of the same molecules observed in this work will allow to derive physical parameters (volume density and temperature) of the streamer and its landing site. 

The presence of the streamer and the redshifted SO component highlight the importance of asymmetric infall for the growth and development of protostars at all evolutionary stages. The high infall rate of this source and the presence of streamers in Class I and II sources suggests that important accretion events of protostars can occur after the Class 0 phase.

\begin{acknowledgements}
\commentstere{The authors would like to thank the anonymous referee for their detailed suggestions, which helped improve the final version of this paper. The authors also thank the IRAM staff at the NOEMA observatory for their invaluable work making these observations possible.} M.T.V. would like to thank L. Testi for his valuable discussions and comments. M.T.V., J.E.P., P.C. and D. S.-C. acknowledge the support by the Max Planck Society. D. S.-C. is supported by an NSF Astronomy and Astrophysics Postdoctoral Fellowship under award AST-2102405. A.F. thanks the Spanish MICINN for funding support from PID2019-106235GB-I00. D.S. acknowledges financial support by the Deutsche Forschungsgemeinschaft through SPP 1833:
``Building a Habitable Earth'' (SE 1962/6-1). I.J.-S. has received partial support from the Spanish State Research Agency (AEI) through project number PID2019-105552RB-C41. \commentstere{N.C. acknowledges funding from  the European Research Council (ERC) via the ERC Synergy Grant \textsl{ECOGAL} (grant 855130). M.T. acknowledges support from project PID2019-108765GB-I00 funded by MCIN/ AEI /10.13039/501100011033. This work is based on observations carried out under project number L19MB with the IRAM NOEMA Interferometer. IRAM is supported by INSU/CNRS (France), MPG (Germany) and IGN (Spain). This research made use of Astropy,\footnote{http://www.astropy.org} a community-developed core Python package for Astronomy \citep{astropy:2013, astropy:2018}.}
\end{acknowledgements}

%
\bibliographystyle{aa} 
\bibliography{main} 

\begin{appendix}

\section{Continuum at 220 GHz \label{sec:cont}}

\begin{figure}[!hbtp]
     \centering
     \includegraphics[width=0.45\textwidth]{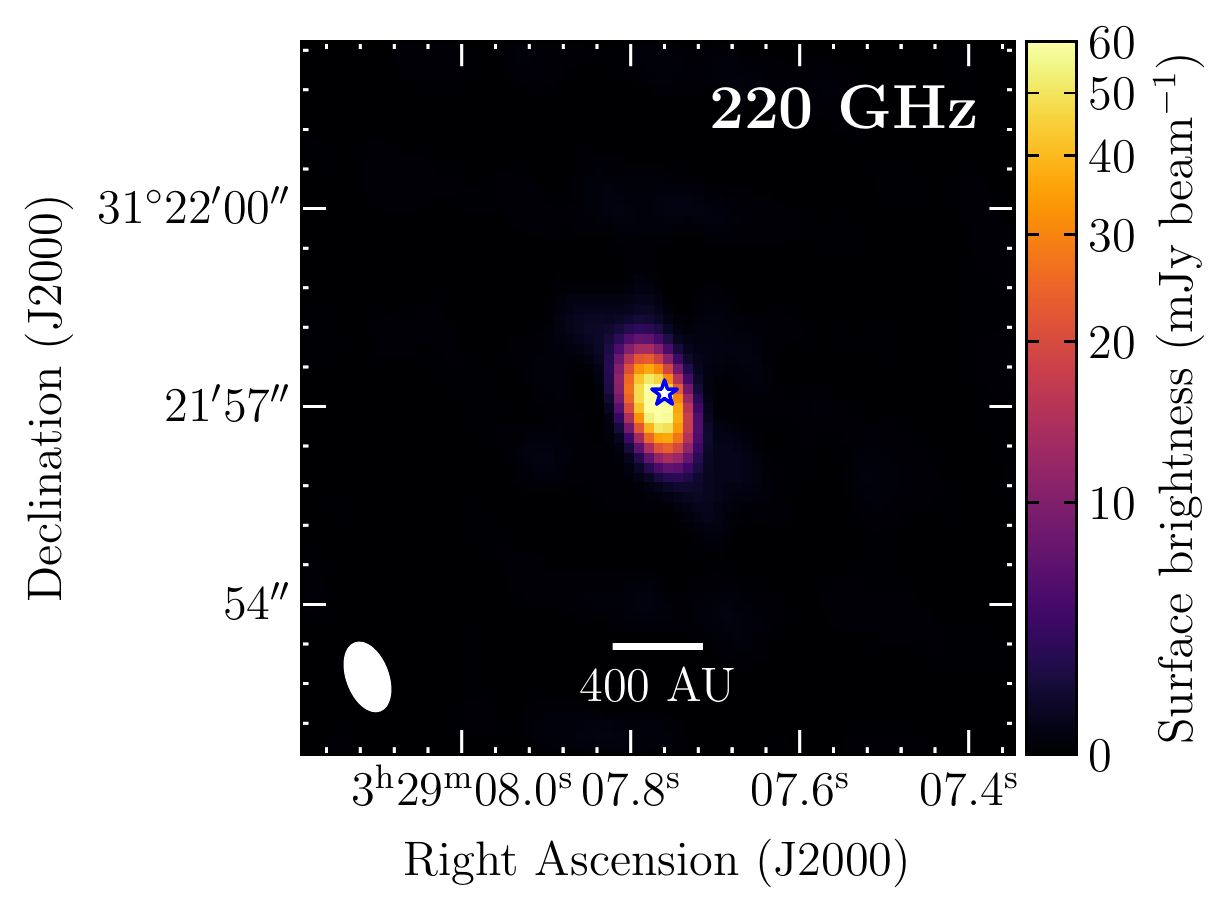}
     \caption{\label{fig:cont} Continuum image at 220 GHz (1.3 mm) of Per-emb-50 obtained with NOEMA. The continuum is done only with the LI continuum window of the observations. The blue star represents the position of Per-emb-50. The white ellipse in the lower left corner represents the beam size.}
\end{figure}

Figure \ref{fig:cont} shows the continuum imageat 1.3 mm (220 GHz) resulting from the LI continuum window of our dataset. The noise level of this image is 0.2 m\Jyb. 

\section{Gaussian components fitting \label{ap:gaussfit}}

We fit a single Gaussian component to all the spectra in the H$_2$CO(3$_{0,3}$--2$_{0,2}$) and the C$^{18}$O(2--1) cubes, using the Python \texttt{pyspeckit} library \citep{ginsburg2011}. We leave out of the analysis all spectra with a peak signal to noise ratio lower than 4. After fitting, we select for further analysis the fitted spectra that meet all of the following requirements:
\begin{itemize}
    \item the parameter uncertainties are all smaller than 50\%,
    \item the Gaussian component has a central velocity in the observed emission velocity range (between 5.5 and 9.5 \kms for H$_2$CO and C$^{18}$O), and
    \item the fitted amplitude has $\mathrm{S/N}>4$.
\end{itemize}

The results of the fit for H$_2$CO(3$_{0,3}$--2$_{0,2}$) are in Fig.~\ref{fig:H2COfit} and for C$^{18}$O(2--1) in Fig.~\ref{fig:C18Ofit}.

We fit one, two and three Gaussian components to the SO spectra near the protostar using the same criteria as above, similar to the multifit approach by \cite{Sokolov2019}. After fitting, we keep the pixels where for each Gaussian, all of the above criteria are met, except for the central velocity, where the emission range changes from $5.5-9.5$ \kms to $-1.0-14.0$ \kms. 

We use the Akaike Information Criterion (AIC) to decide whether one, two or three Gaussian components reproduce best each SO spectra. This criterion uses the AIC value $AIC$ to determine which model minimizes information loss:
\begin{equation}
    AIC = 2k + \chi^2 + C,
\end{equation}
where $k$ is related to the number of free parameters of the model (see below), $\chi^2$ is the classical chi-squared statistic and C is a constant defined by the number of independent channels and the uncertainties \citep{Choudhury2020}. For $k$, each Gaussian component has three free parameters, so $k=3g$, where $g$ is the number of Gaussian components in each model. For $C$, we assume that each channel in the spectra has a constant normal error, which corresponds to the rms of the SO cube, and we use the same data to test the three models, C is the same for all models and does not play a role in choosing the best model, so we set $C=0$. The fit with the lowest AIC value is the preferred one for each spectrum.

We evaluate the probability that the model with the minimum information loss is a considerable improvement from the other two models for each spectrum. The difference between the minimum AIC, $AIC_{min}$ (which comes from the "best" model) and the AIC value of model i, $AIC_{i}$, is proportional to the probability that model i is as good as the minimum to minimize information loss as:

\begin{equation}
    P \propto \exp \Big(\frac{AIC_{min}-AIC_{i}}{2}\Big).
\end{equation}

For SO($5_5-4_4$), all of the fitted spectra have less than 5\% probability than the competing models minimize the information loss better than the model with minimum $AIC$. This means that, for those spectra that are well fitted by three Gaussians, the improvement from two Gaussians is significant. The same can be said for the improvement in those spectra that are best fitted with two Gaussians instead of only one. Therefore, we conclude that each spectra is well described by one, two or three Gaussian components, depending on each case. 

\begin{figure*}[htbp]
     \centering
     \includegraphics[width=0.30\textwidth]{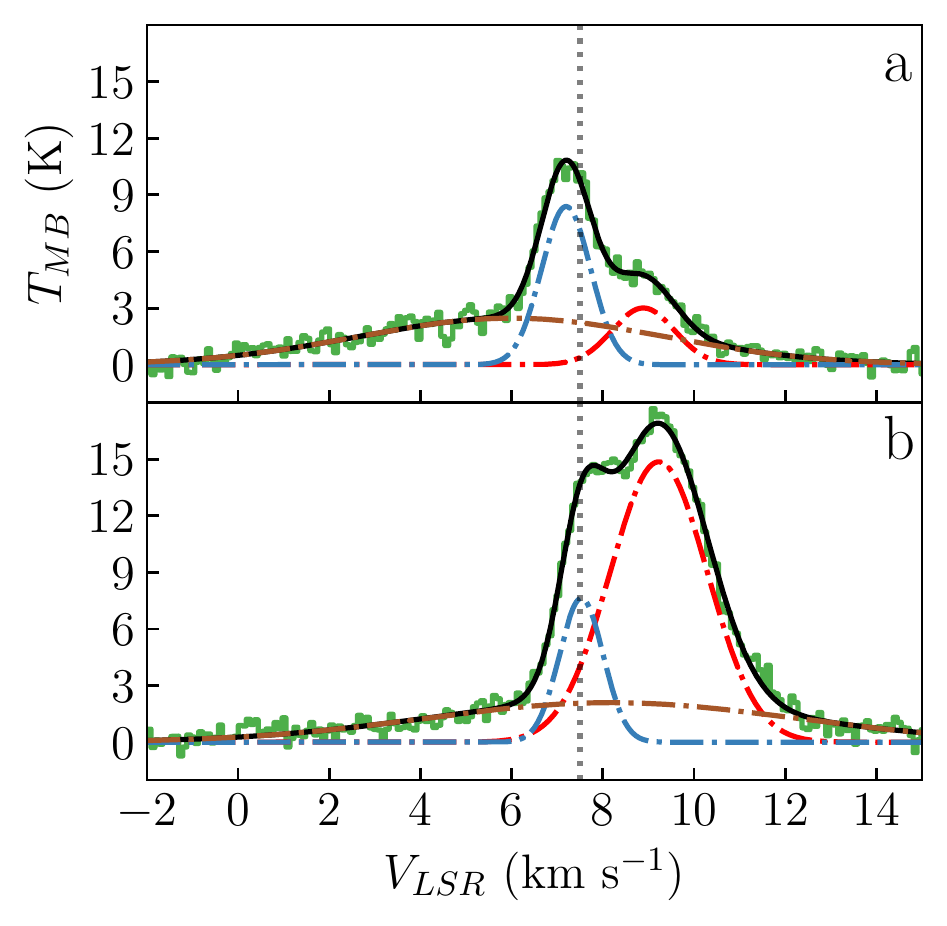}
     \includegraphics[width=0.39\textwidth]{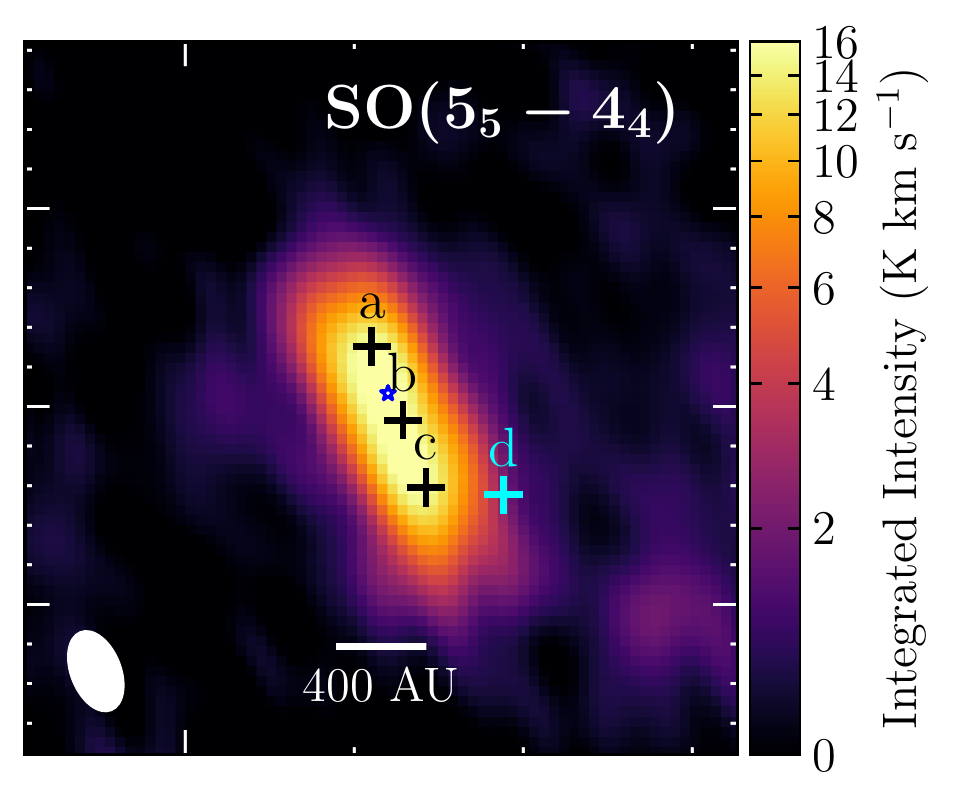}
     \includegraphics[width=0.30\textwidth]{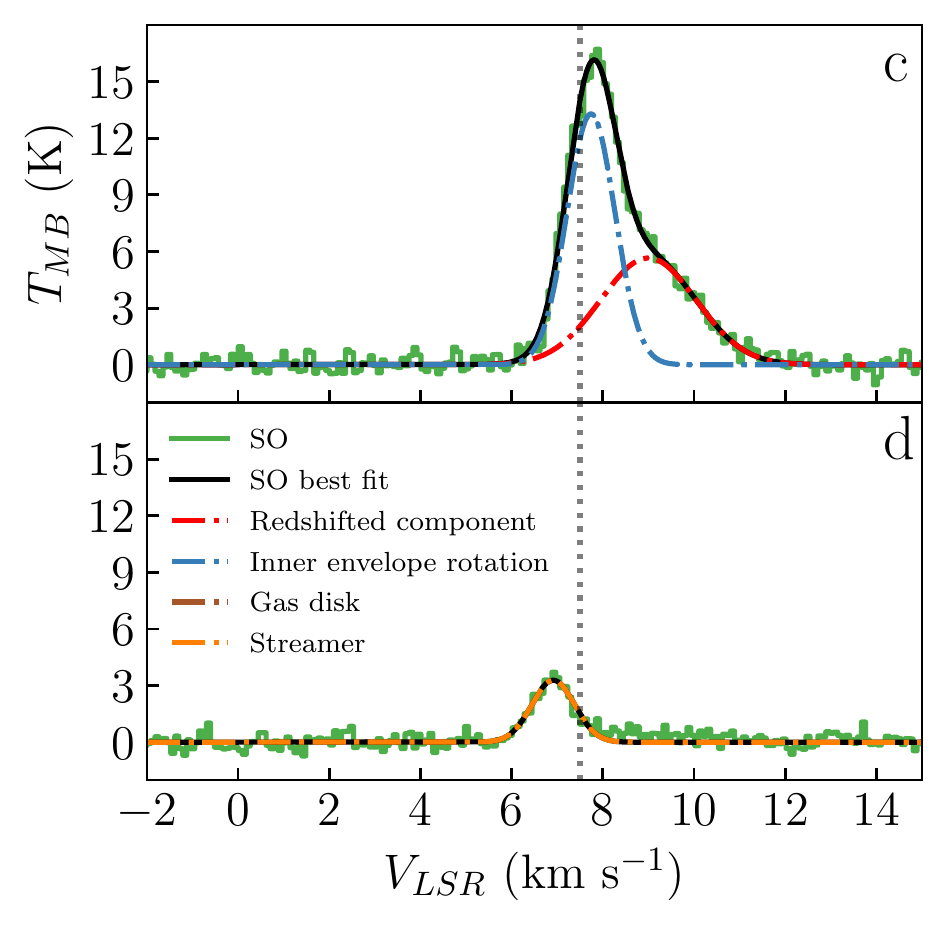}
    \caption{\label{fig:SOspectra} Sample spectra of SO in 4 selected locations in Per-emb-50's inner envelope and disk region. \textbf{Left and right:} Beam averaged SO spectra (green line) at positions a to d. The black solid curve represents the best fit Gaussian function. The dashed-dotted lines represent the individual Gaussians that, added together, correspond to the best fit function. Each individual color in the dashed-dotted lines represents one of the kinematic components found in Sect. \ref{sec:SOdecomposition}: blue corresponds to the inner envelope rotation, red matches the redshifted component, brown corresponds to the gas disk and orange matches the streamer. \textbf{Middle:} Integrated intensity map of SO between -1 and 14, as shown in Fig.~\ref{fig:images}, zoomed into the inner 1000 au closest to the protostar. Crosses (black and cyan) mark the locations of each spectra, labeled a to d from highest to lowest right ascension. }
\end{figure*}

Figure \ref{fig:SOspectra} shows four spectra fitted with either one, two or three Gaussians. 

\section{Envelope mass calculation\label{ap:envelopemass}}

We obtained the envelope mass upper and lower limits using the flux in  our continuum obtained with NOEMA (see Appendix \ref{sec:cont}) and the Bolocam 1.1 mm image from \cite{Enoch2006}. First, we obtain the flux in the Bolocam 1.1 mm continuum within a beam-sized aperture ($\theta_{FWHM}=31\arcsec$), centered at the location Per-emb-50, $F_{\text{Bolocam}} = 324 \pm 46$ mJy, together with the peak value within this aperture $I_{\text{Bolocam}} = 573 \pm 55$ m\Jyb. Then, we obtain the total flux and peak value within the primary beam of the continuum obtained with NOEMA ($22\arcsec$), $F_{\text{NOEMA}} = 89 \pm 2$ mJy and $I_{\text{NOEMA}} = 72.9 \pm 1$ m\Jyb, respectively. We assume that the NOEMA continuum contains disk emission only, as it does not contain zero-spacing information, whereas the Bolocam 1.1 mm image includes emission from the disk and envelope. We subtract the flux in the NOEMA continuum to the flux obtained from Bolocam, thus obtaining the flux of the envelope only $\Delta S_{1\text{mm}} = S_{\text{Bolocam}} - S_{\text{NOEMA}}$, and use Equation 4 of \cite{Enoch2009} to calculate the envelope mass:

\begin{equation}
    M_{env} = \frac{D^2 \Delta S_{1\text{mm}}}{B_{1\text{mm}}(T_D)\kappa_{1\text{mm}}}.
\end{equation}
We assume that the continuum at 1 mm consists of optically thin emission and use  $\kappa_{1mm}=0.0114$ cm$^{2}$ g$^{-1}$, $T_D=15$ K as stated in \cite{Enoch2009}, and a distance $D=293$ pc \citep{Ortiz-Leon2018}. Using the flux difference we obtain an envelope mass of 0.18 \Msun and with the peak difference, 0.39 \Msun.

\section{Determination of column density \label{ap:columndens}}
We first obtain the integrated intensity map of the primary beam corrected C$^{18}$O(2 --1) emission in the spatial region where the streamer is defined for the streamline model (see Fig.~\ref{fig:C18Ofit}). We integrate the map between 5.5 and 9.5 \kms. This velocity range covers the spectral emission of the streamer in C$^{18}$O(2 --1) completely. Then, we calculate the total column density of the C$^{18}$O molecule using equation 80 of \cite{Mangum2015} in each pixel of the integrated intensity map. We use a line strength $S=\frac{J^2}{J(2J+1)} = 2/5$ in relation to the dipole moment of the C$^{18}$O molecule $\mu=0.11079$ Debye $=1.1079\times10^{-19}$ esu cm, the rotor rotation constant for C$^{18}$O $B_0 = 54891.420$ MHz, the upper state energy for the C$^{18}$O(2 --1) transition $E_u=15.81$ K, and the degeneracy of the C$^{18}$O (2 -- 1) transition $g_J = 2J+1 = 5$. We assume a beam filling factor $f=1$, as emission is resolved. The resulting equation for $N(\mathrm{C}^{18}\mathrm{O})$ in cm$^{-2}$, $T_{ex}$ in K and $\int T_R\, dv$ in K \kms is

\begin{multline}
    N(\mathrm{C}^{18}\mathrm{O}) =  1.63 \times 10^{15} \frac{Q_{rot}(B_0, T_{ex})}{5} \frac{\exp(\frac{15.81}{T_{ex}})}{\exp(\frac{10.54}{T_{ex}})-1} \\
    \times \frac{\int T_R\, dv}{J_{\nu}(T_{ex})-J_{\nu}(T_{bg})},
\end{multline}
where
\begin{equation}
    Q_{rot} = \frac{k_B T_{ex}}{h B_0} + \frac{1}{3}
\end{equation}
is the first order Taylor approximation of the partition function of a rigid-rotor diatomic molecule, and 
\begin{equation}
    J_{\nu}(T) = \frac{\frac{h\nu}{k_B }}{\exp(\frac{h\nu}{k_B T})-1}
\end{equation}
is the Rayleigh-Jeans equivalent temperature in K. We use $T_{bg}=2.7$ K and $\nu=219.560$ GHz (the frequency of the C$^{18}$O(2 --1) line). We use a constant $T_{ex}=15\pm5$ K. 

\section{SO$_2$ spectra and image \label{sec:SO2}}

\begin{figure*}[htbp]
     \centering
     \includegraphics[width=0.29\textwidth]{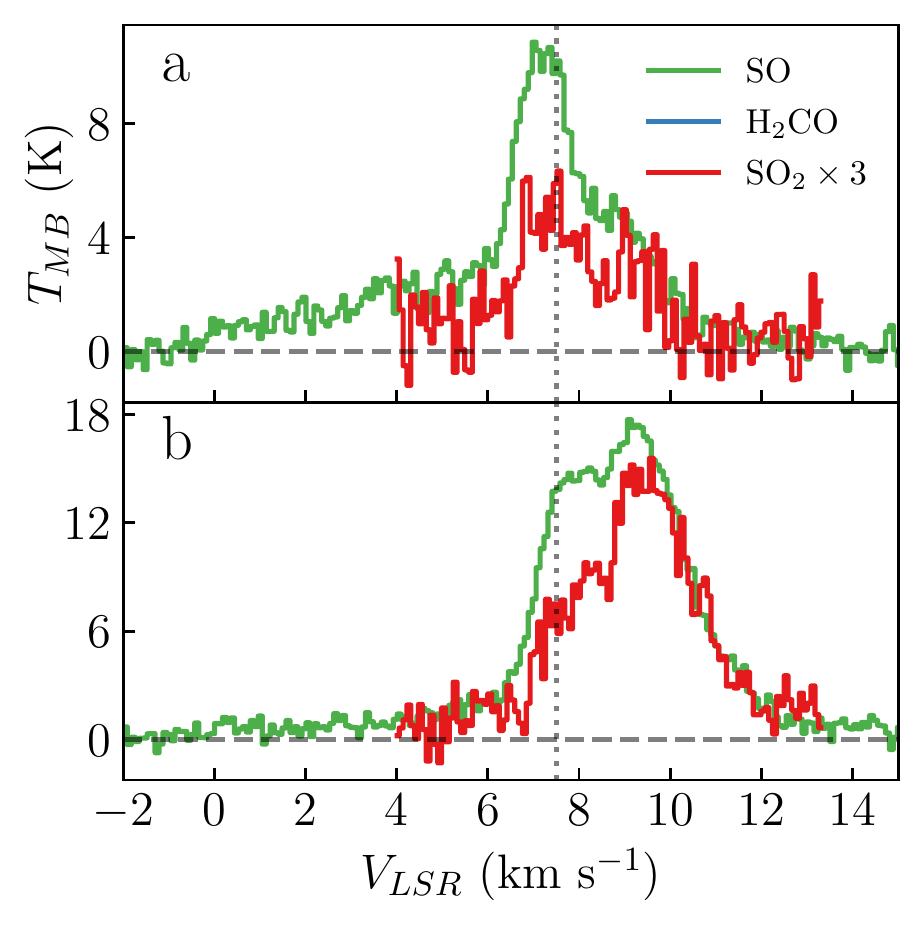}
     \includegraphics[width=0.41\textwidth]{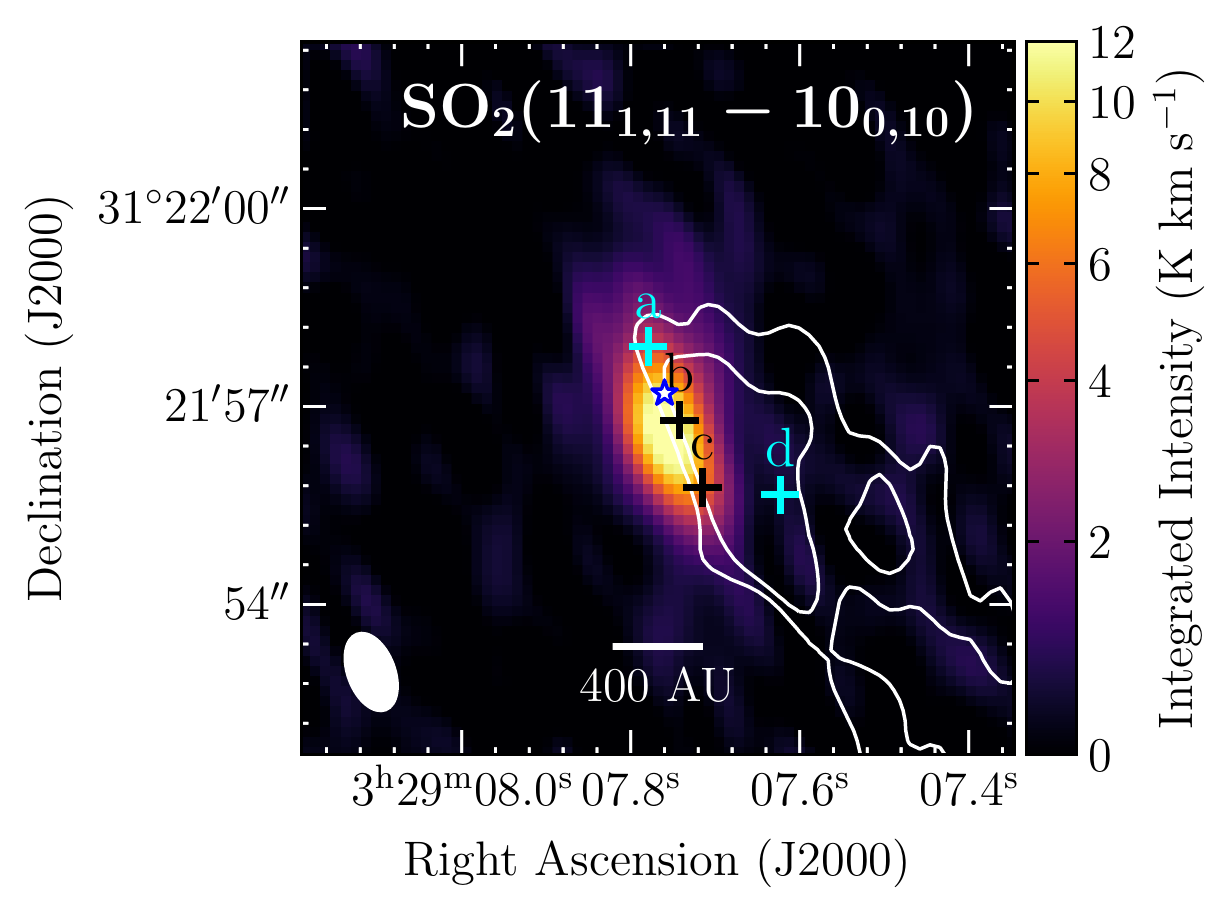}
     \includegraphics[width=0.29\textwidth]{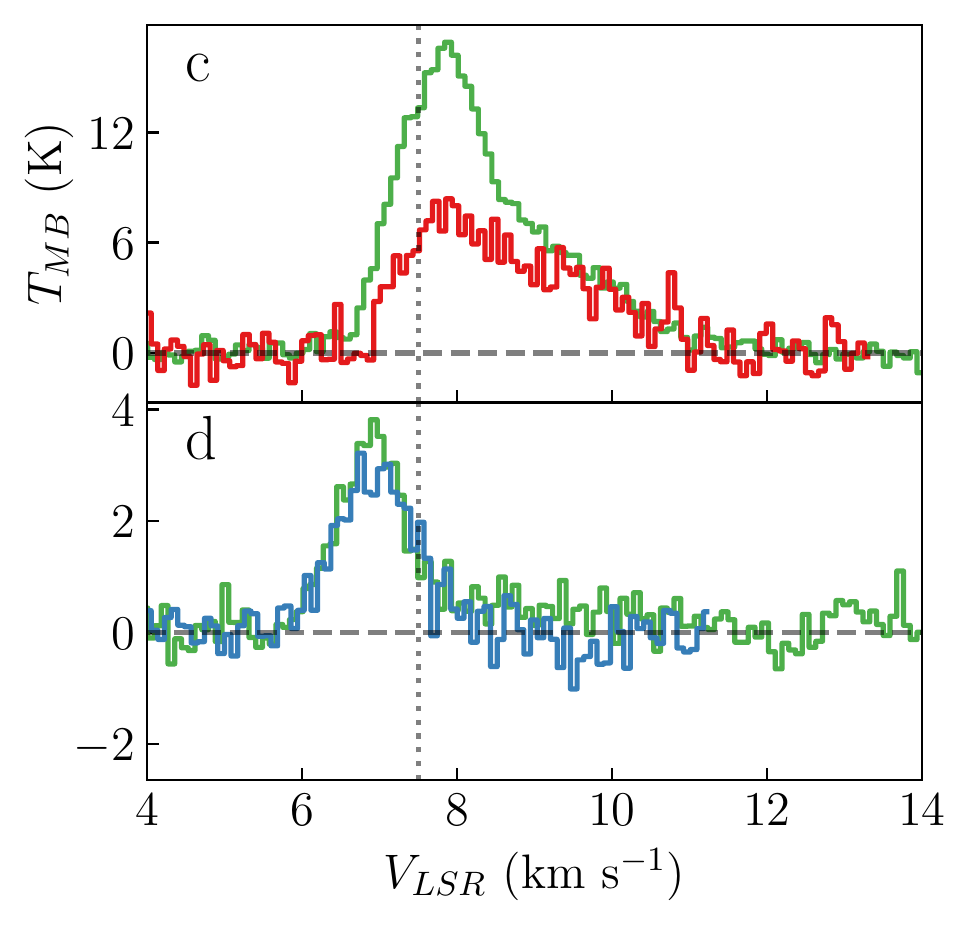}
     \caption{\label{fig:SO2withspectra} Velocity integrated SO$_2$(11$_{1,11}$ -- 10$_{0,10}$) together with sample spectra of SO$_2$, SO and H$_2$CO at positions a to d. \textbf{Left and Right:} Beam averaged spectra at positions a to d. Each color represents the spectrum of a molecule: blue corresponds to H$_2$CO, green to SO and red to SO$_2$. The dotted vertical line marks the central velocity of the protostar $V_{LSR} = 7.5$ \kms. The dashed horizontal line represents the 0 K level. If a molecule does not appear at a panel, its because emission from that molecule has $\mathrm{S/N}<5$. \textbf{Middle:} Velocity integrated image of the SO$_2$(11$_{1,11}$ -- 10$_{0, 10}$) molecular transition. The crosses represent the places where each spectra was extracted, labeled from a to d. The white contours represent the 3 and 5 times the rms of the integrated map contours of H$_2$CO (0.25 K \kms). The white ellipse in the bottom left corner represents the beam size. The blue star marks the position of Per-emb-50.}
\end{figure*}

Figure \ref{fig:SO2withspectra} shows the integrated intensity map of SO$_2$(11$_{1,11}$ -- 10$_{0,10}$) between 5 and 12 \kms and to the left and right, spectra of SO, SO$_2$ and H$_2$CO in the same selected positions as in Fig.~\ref{fig:SOspectra}. 

\section{SO decomposition \label{ap:SOdecomp-results}}

\begin{figure*}
    \centering
    \includegraphics[width=0.45\textwidth]{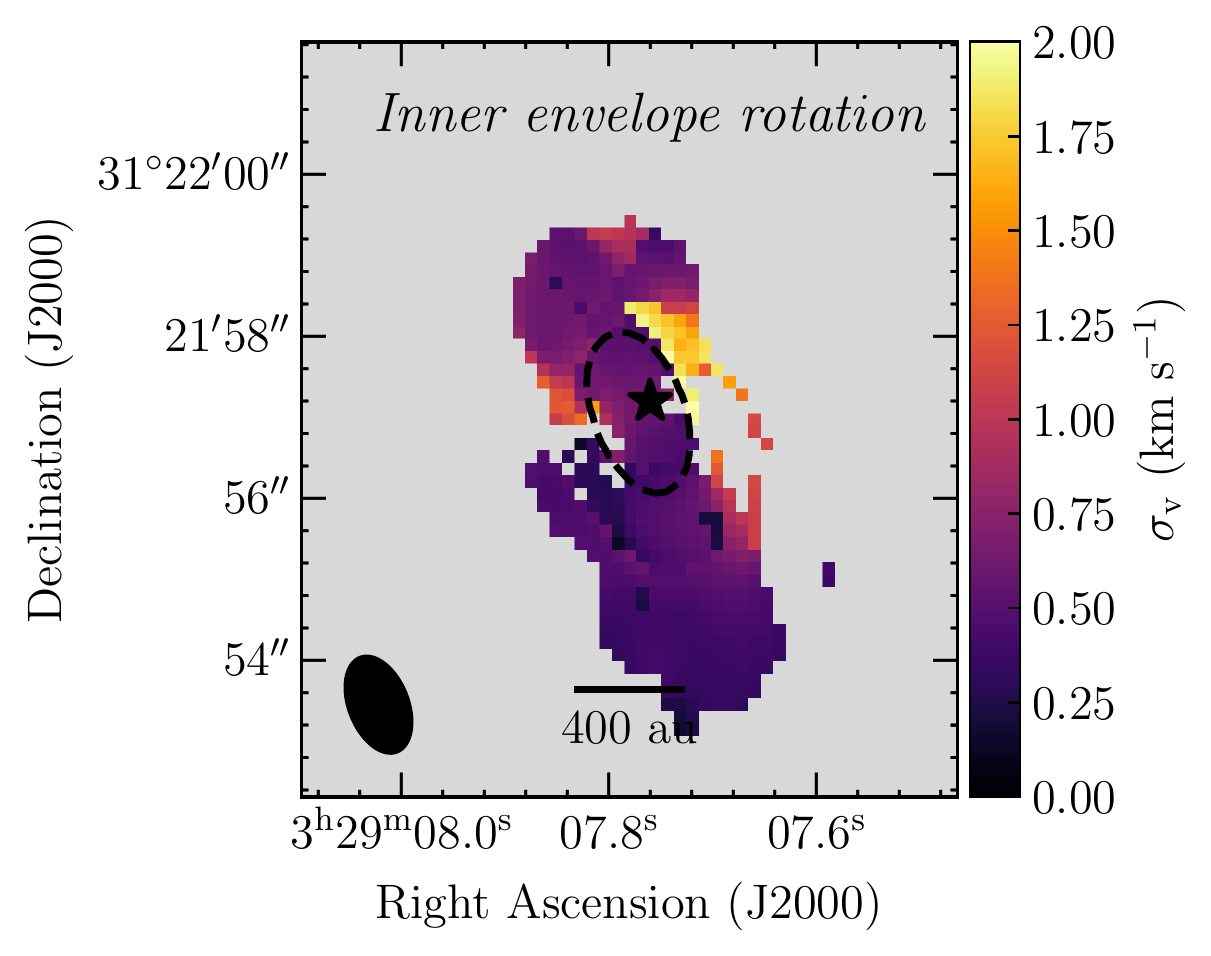}
    \includegraphics[width=0.45\textwidth]{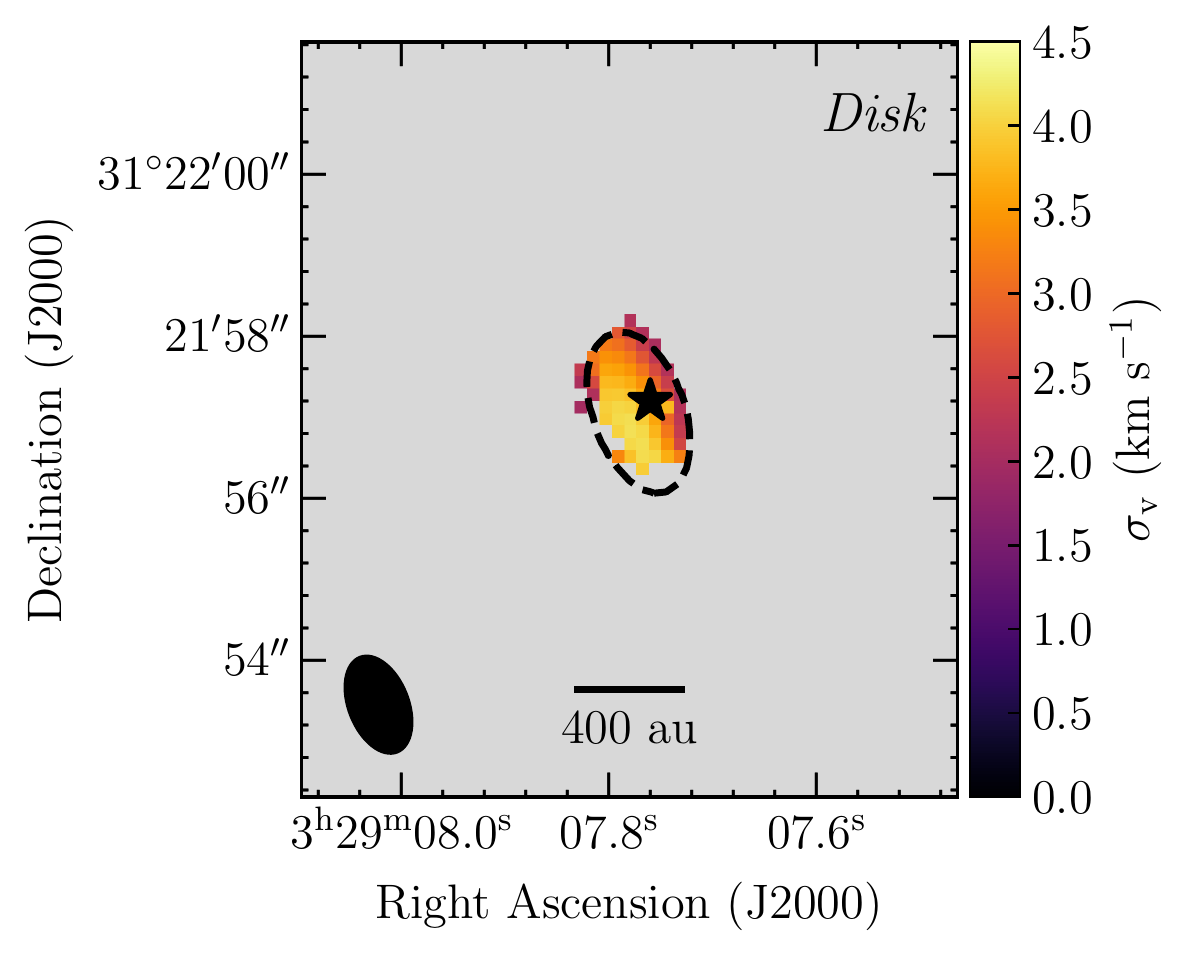}
    \includegraphics[width=0.45\textwidth]{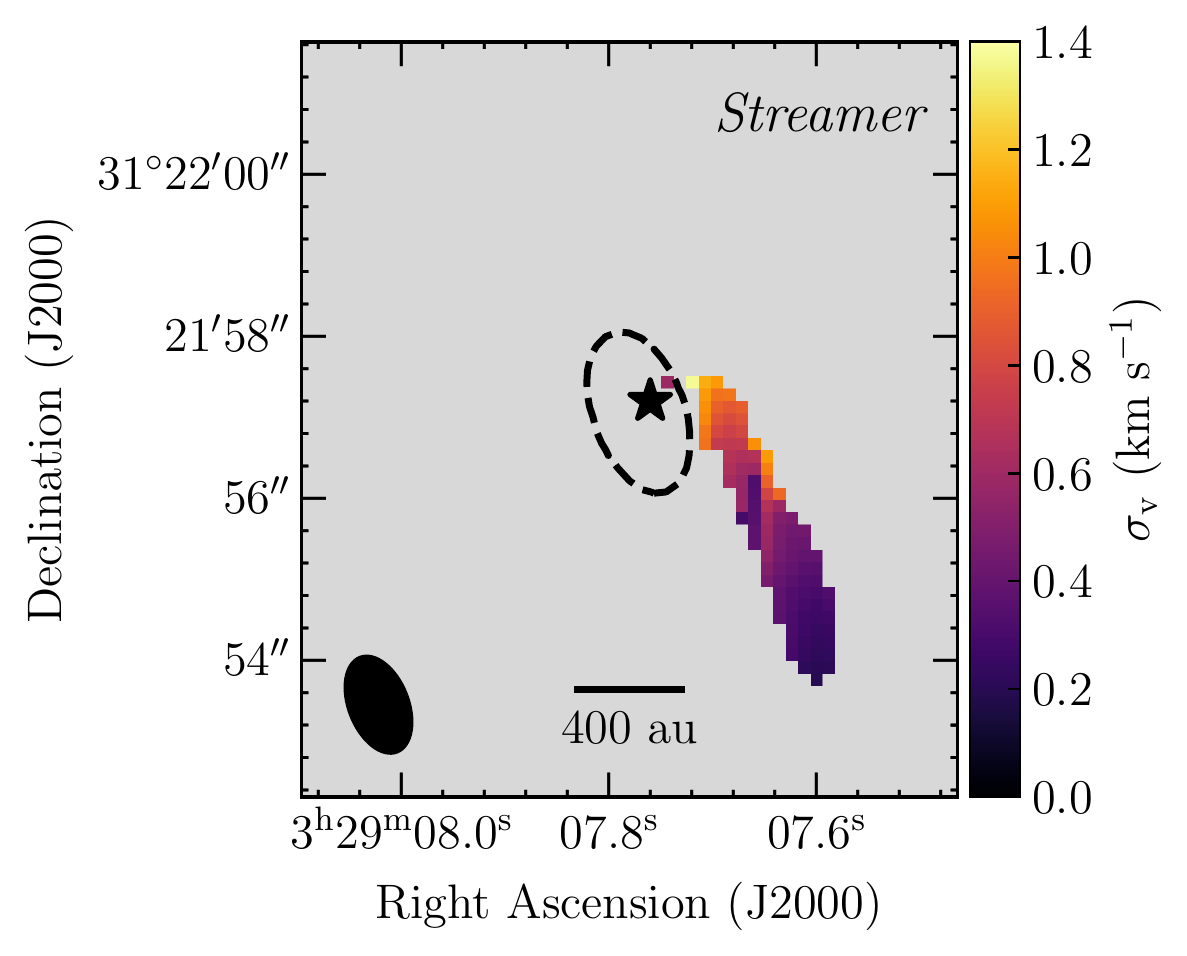}
    \includegraphics[width=0.45\textwidth]{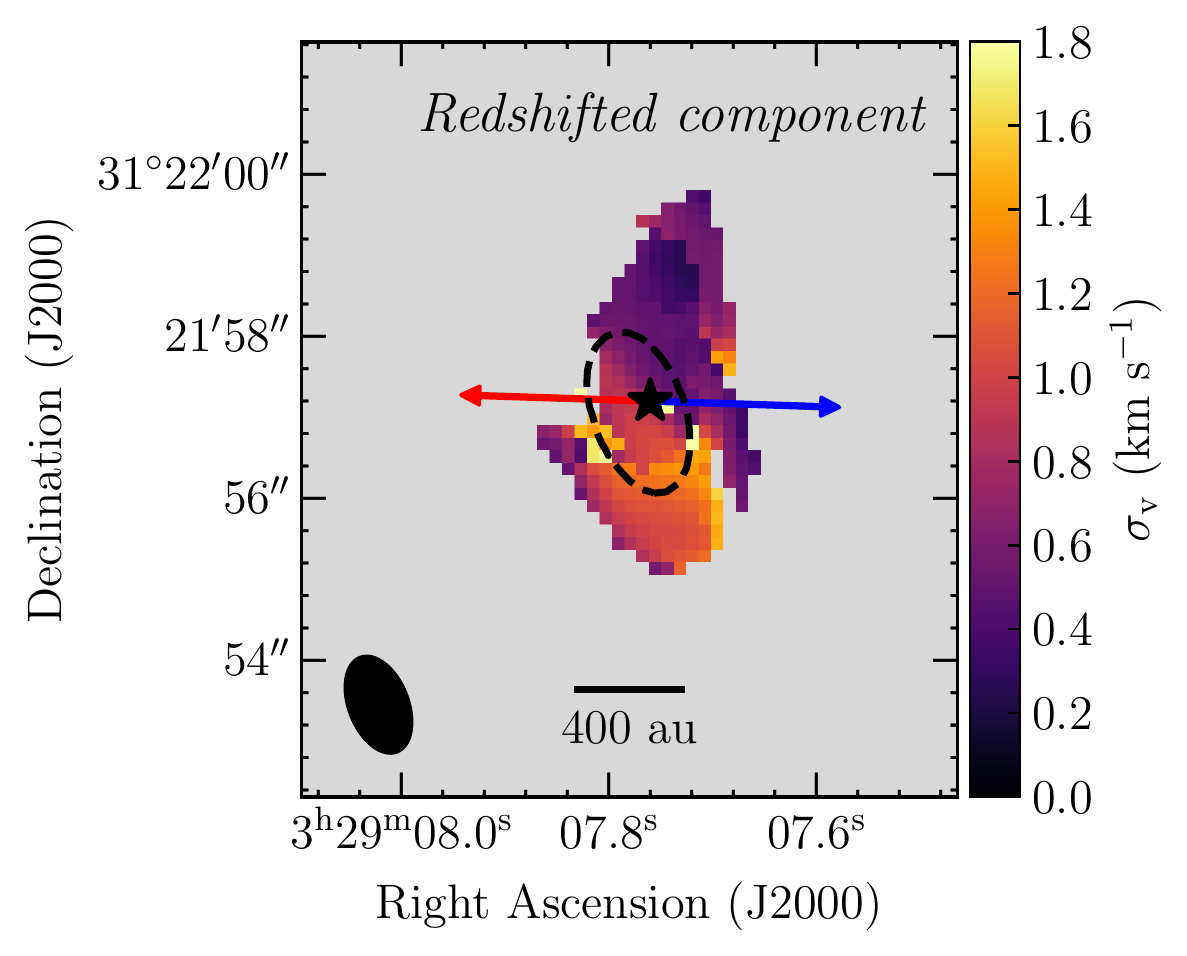}
    \caption{$\sigma_{\mathrm{v}}$ of the \textit{inner envelope rotation}, \textit{disk}, \textit{streamer} and \textit{redshifted} components in SO($5_5-4_4$) found in Sect. \ref{sec:SOdecomposition}. The dashed contour represents the 220 GHz continuum emission at the 7 m\Jyb level. The black ellipse in the lower left corners represent the beam size.}
    \label{fig:aux-SOsigmav}
\end{figure*}

Figure \ref{fig:aux-SOsigmav} shows the velocity dispersion $\sigma_{\mathrm{v}}$ of each kinematic element found in Sect. \ref{sec:SOdecomposition} (see Fig.~\ref{fig:SOcomponents}) through the Gaussian fitting described in Appendix \ref{ap:gaussfit}. Note that all images have different colorscales.

\end{appendix}
\end{document}